\documentclass[final]{ias2}

\usepackage{graphicx} 
\usepackage{multirow}
\usepackage{array} 
\usepackage{xspace}

\usepackage{hyperref} 
\mathchardef\Upsilon="7107
\def\Y#1S{\ensuremath{\Upsilon{(#1S)}}\xspace}

\newcommand{\mtau}{\ensuremath{m_\tau}\xspace}
\newcommand{\mZ}{\ensuremath{M_Z}\xspace}
\newcommand{\as}{\ensuremath{\alpha_{\scriptscriptstyle S}}\xspace}
\newcommand{\asTau}{\ensuremath{\as(\mtau)}\xspace}

\newcommand{\Kbar    }{\kern 0.2em\overline{\kern -0.2em K}{}\xspace}
\newcommand{\Dbar    }{\kern 0.2em\overline{\kern -0.2em D}{}\xspace}

\newcommand{\Kz      }{\ensuremath{K^0}\xspace}
\newcommand{\Kzb     }{\ensuremath{\Kbar^0}\xspace}
\newcommand{\KzKzb   }{\ensuremath{\Kz \kern -0.16em \Kzb}\xspace}
\newcommand{\Kp      }{\ensuremath{K^+}\xspace}
\newcommand{\Km      }{\ensuremath{K^-}\xspace}

\newcommand{\KpKm    }{\ensuremath{\Kp \kern -0.16em \Km}\xspace}
\newcommand{\KS      }{\ensuremath{K^0_{\scriptscriptstyle S}}\xspace} 
\newcommand{\KL      }{\ensuremath{K^0_{\scriptscriptstyle L}}\xspace}

\newcommand{\e}{\varepsilon}

\newcommand{\nub}{\ensuremath{\overline{\nu}}\xspace}

\newcommand{\BR}{\ensuremath{{\cal B}}\xspace}

\newcommand{\ee}{\ensuremath{e^+e^-}\xspace}

\newcommand{\pp}{\ensuremath{\pi^+\pi^-}\xspace}

\newcommand{\tev}{\ensuremath{\mathrm{Te\kern -0.1em V}}\xspace}
\newcommand{\gev}{\ensuremath{\mathrm{Ge\kern -0.1em V}}\xspace}
\newcommand{\mev}{\ensuremath{\mathrm{Me\kern -0.1em V}}\xspace}
\newcommand{\kev}{\ensuremath{\mathrm{ke\kern -0.1em V}}\xspace}
\newcommand{\ev}{\ensuremath{\mathrm{e\kern -0.1em V}}\xspace}
\newcommand{\gevc}{\ensuremath{{\mathrm{Ge\kern -0.1em V\!/}c}}\xspace}
\newcommand{\mevc}{\ensuremath{{\mathrm{Me\kern -0.1em V\!/}c}}\xspace}
\newcommand{\gevcc}{\ensuremath{{\mathrm{Ge\kern -0.1em V\!/}c^2}}\xspace}
\newcommand{\mevcc}{\ensuremath{{\mathrm{Me\kern -0.1em V\!/}c^2}}\xspace}

\newcommand{\bei}{\begin{itemize}}
\newcommand{\eei}{\end{itemize}}
\newcommand{\ben}{\begin{enumerate}}
\newcommand{\een}{\end{enumerate}}
\newcommand{\beq}{\begin{equation}}
\newcommand{\eeq}{\end{equation}}
\newcommand{\beqn}{\begin{eqnarray}}
\newcommand{\eeqn}{\end{eqnarray}}
\newcommand{\beqns}{\begin{eqnarray*}}
\newcommand{\eeqns}{\end{eqnarray*}}

\newcommand{\amu}{\ensuremath{a_\mu}\xspace}
\newcommand{\amuhad}{\ensuremath{\amu^{\rm had}}\xspace}
\newcommand{\amuhadLO}{\ensuremath{\amu^{\rm had,LO}}\xspace}

\newcommand{\amuhadHO}{\ensuremath{\amu^{\rm had,NLO}}\xspace}
\newcommand{\amuhadHODisp}{\ensuremath{\amuhadHO[{\rm Disp}]}\xspace}
\newcommand{\amuhadHOLBLS}{\ensuremath{\amuhadHO[{\rm LBL}]}\xspace}
\newcommand\ie{{i.e.}\xspace} 
\newcommand\eg{{e.g.}\xspace}

\newcommand\cf{{cf.}\xspace}

\def\@citex[#1]#2{\if@filesw\immediate\write\@auxout{\string\citation{#2}}\fi
  \@tempcnta\z@\@tempcntb\m@ne\def\@citea{}\@cite{\@for\@citeb:=#2\do
    {\@ifundefined
       {b@\@citeb}{\@citeo\@tempcntb\m@ne\@citea
        \def\@citea{,\penalty\@m\ }{\bf ?}\@warning
       {Citation `\@citeb' on page \thepage \space undefined}}%
    {\setbox\z@\hbox{\global\@tempcntc0\csname b@\@citeb\endcsname\relax}%
     \ifnum\@tempcntc=\z@ \@citeo\@tempcntb\m@ne
       \@citea\def\@citea{,\penalty\@m}
       \hbox{\csname b@\@citeb\endcsname}%
     \else
      \advance\@tempcntb\@ne
      \ifnum\@tempcntb=\@tempcntc
      \else\advance\@tempcntb\m@ne\@citeo
      \@tempcnta\@tempcntc\@tempcntb\@tempcntc\fi\fi}}\@citeo}{#1}}

\def\@citeo{\ifnum\@tempcnta>\@tempcntb\else\@citea
  \def\@citea{,\penalty\@m}%
  \ifnum\@tempcnta=\@tempcntb\the\@tempcnta\else
   {\advance\@tempcnta\@ne\ifnum\@tempcnta=\@tempcntb \else
\def\@citea{--}\fi
    \advance\@tempcnta\m@ne\the\@tempcnta\@citea\the\@tempcntb}\fi\fi}
\newenvironment{myquote}
               {\list{}{\leftmargin0cm\indent}%
                \item\relax}
               {\endlist}
\newcommand\allFontSize{\footnotesize}
\newcommand\detailsSize{\allFontSize}
{\begin{myquote}\detailsSize}{\end{myquote}}

\RequirePackage{xspace}

\def\CP                {{\em CP}\xspace}

\newcommand{\mue}{\ensuremath{\mbox{$\mu$--$e$}}\xspace}

\def\Bbar    {\kern 0.18em\overline{\kern -0.18em B}{}\xspace}

\def\Bz      {\ensuremath{B^0}\xspace}
\def\Bzb     {\ensuremath{\Bbar^0}\xspace}
\def\Bu      {\ensuremath{B^+}\xspace}
\def\Bub     {\ensuremath{B^-}\xspace}

\def\BpBm    {\ensuremath{\Bu {\kern -0.16em \Bub}}\xspace}
\def\Bs      {\ensuremath{B^0_{s}}\xspace}
\def\Bsb     {\ensuremath{\Bbar^0_{s}}\xspace}

\def\BzBzb   {\ensuremath{\Bz {\kern -0.16em \Bzb}}\xspace}
\def\BszBszb {\ensuremath{\Bs {\kern -0.16em \Bsb}}\xspace}

\newcommand{\tanb}{\ensuremath{\tan\!\beta}\xspace}

\begin{document}

\markboth{Charged-Lepton Flavour Physics}{Andreas Hoecker}

\title{Charged-Lepton Flavour Physics}

\author[sin]{Andreas Hoecker} 
\email{andreas.hoecker@cern.ch}
\address[sin]{CERN, Switzerland}

\begin{abstract}
This writeup of a talk at the 2011 Lepton-Photon symposium in Mumbai, India, 
summarises recent results 
in the charged-lepton flavour sector. I review searches for charged-lepton 
flavour violation, lepton electric dipole moments and flavour-conserving \CP violation.
I also discuss recent progress in $\tau$-lepton physics and in the Standard Model 
prediction of the muon anomalous magnetic moment. 
\end{abstract}


 
\maketitle


\section{Introduction}

The instability of the scalar sector of the Standard Model (SM) with 
respect to fermionic and bosonic loop corrections in presence of an
ultraviolet cut-off scale has been the driving motivation for the 
widespread expectation of TeV-scale new physics. Such new physics
is in reach for discovery at the LHC which sets the current experimental energy 
frontier. Clean measurements in the charged-lepton flavour sector can also 
probe new physics at high scales, in fact, at mass scales up to hundreds 
of TeV and above, albeit the interpretation of a measurement in terms
of mass scales is model-dependent. Such measurements often
involve rare processes or small deviations in abundant processes 
(because they are scale suppressed) and hence require large rates and high 
precision to be observable. The experiments performing these measurements
operate at the intensity and precision frontier. Both, the energy and 
intensity/precision frontiers are complementary domains of activity in modern 
particle physics research that must be pursued in common. 

Although some deviations between experiment and Standard Model expectation exist 
in the charged flavour physics sector, none is presently significant enough to 
demonstrate evidence for new physics. It is nevertheless important to follow up 
on them. Just as important are the measurements for which, in our ignorance, 
one would have expected new physics to show up but none has been seen. Indeed, 
many of the models
that have been developed to stabilise the Higgs boson sector, or that suggest
alternatives to it, predict new flavour-changing neutral currents, \CP-violating 
phases, charged-lepton flavour violation, electric dipole moments, anomalous 
magnetic moments, contributions to electroweak precision observables, etc. Their 
non-observation strongly constrains these models, though -- apart from 
very specific cases -- does not exclude them. It does, however, affect their 
naturalness and the searches at the LHC, be they positive or not, will need 
to be scrutinised in view of the results from the precision measurements.

I review in the following searches for charged-lepton flavour violation, 
lepton electric dipole moments and flavour-conserving \CP violation.
I also discuss recent progress in $\tau$-lepton physics and in the Standard 
Model prediction of the muon anomalous magnetic moment.

\section{Charged-lepton flavour violation}

Flavour violation involving charged leptons (LFV) belongs to the class of 
flavour-changing neutral currents (FCNC), which are suppressed at tree level
in the SM where they are mediated by $\gamma$ and $Z^0$ bosons, but arise at 
loop level via weak charged currents mediated by the $W^\pm$ boson. The GIM 
mechanism~\cite{Glashow:1970gm} further suppresses loop-induced FCNC in 
the quark sector, so that FCNC effects are generally small in the SM. Rare
FCNC processes such as $\Bs\to\mu\mu$ or $K^+\to\pi^+\nu\nub$ 
(and many others) are therefore sensitive probes for new physics. The former mode
is currently actively investigated at the LHC~\cite{LHCb:2011ac,PhysRevLett.107.191802},
while the latter channel will be studied by the NA62 experiment that is 
under construction at CERN~\cite{na62}.

Because flavour violation requires mixing between generations, charged LFV exactly vanishes 
in the SM for massless neutrinos. Extending the SM to include neutrino masses
induces charged LFV via chirality flipping dipole amplitudes, which are however 
proportional to the fourth power in the ratio of neutrino mass splitting to $W$
mass, giving, \eg, for the LFV decay $\mu\to e\gamma$ a branching fraction of 
roughly $10^{-54}$~\cite{Marciano:2008zz}, 
depending on the neutrino mixing angle $\theta_{13}$. This is an unobservably tiny
branching fraction so that the search for charged LFV probes new physics without 
SM contamination.

\begin{figure}[t]
\begin{center}
\includegraphics[width=0.7\columnwidth]{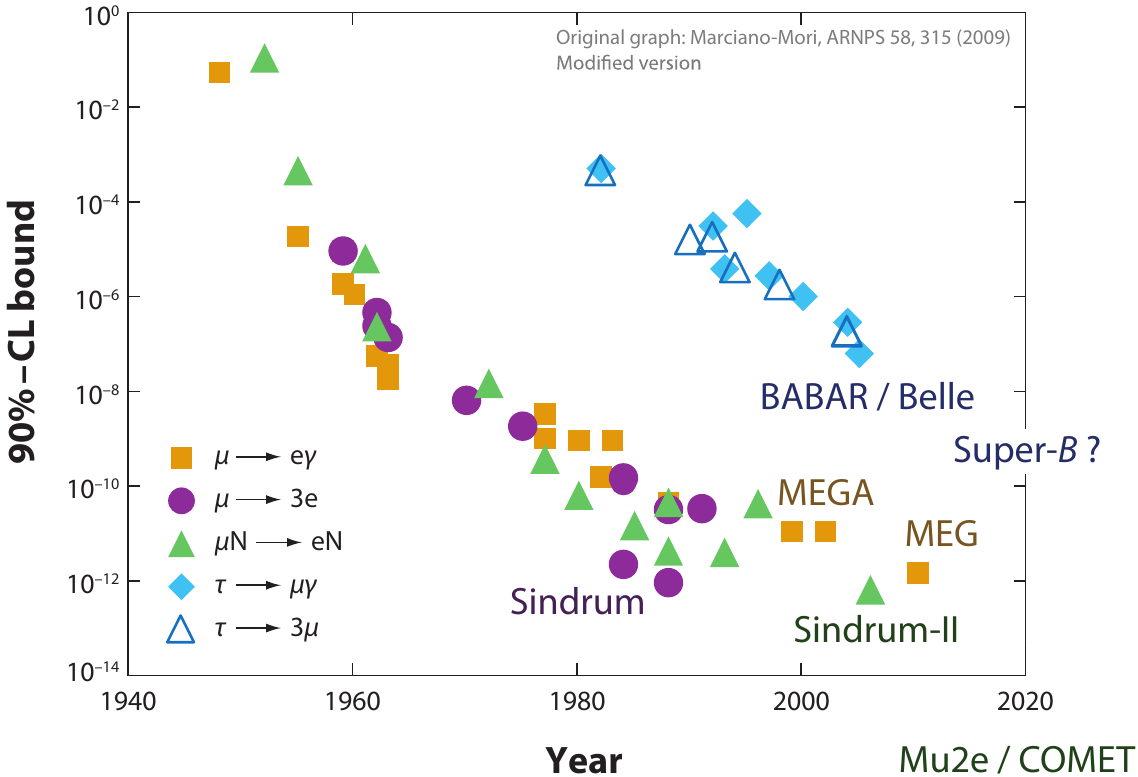}
\end{center}
\vspace{-0.3cm}
\caption{\label{fig:lfv-history}
  History of searches for selected lepton flavour violating processes.
  Shown are 90\% CL upper limits, and the experiments setting
  the best current limits and future prospectives for LFV searches in 
  $\tau$ decays and \mue conversion are indicated. 
  This graph has been modified from~\cite{Marciano:2008zz}.}
\end{figure}

Experimentally, no evidence for charged LFV has been found so far. It is searched for 
in a variety of modes including the neutrinoless decays of a heavy lepton into a light 
one under emission of a radiative photon, or of a heavy lepton into three light ones. 
Using muonic atoms it is also possible to look for \mue conversion in the electromagnetic 
field of the nucleus. Finally, $\tau$ leptons provide a profuse field of LFV searches 
with 48 different final states studied so far (see~\cite{Asner:2010qj} for a recent 
summary). A chronological overview of LFV limits is drawn in Fig.~\ref{fig:lfv-history}.
It witnesses the many orders of magnitude improvement in the sensitivity obtained 
during half a century of LFV experiments. The tightest absolute limits on LFV effects 
are obtained in $\mu$ decays and \mue conversion experiments. However, because 
different new physics phenomena induce different LFV effects, a quantitative 
comparison between the limits is model-dependent.

The absence of charged LFV had important consequences in the early days of particle
physics when the concept of fermion generations was developed. Non-discovery of 
$\mu\to e\gamma$ and $\tau\to\mu\gamma$ established that $\mu$ and $\tau$ were indeed 
new elementary leptons, as opposed to excited states of composite lighter leptons. 
In analogy to the GIM mechanism, the absence of $\mu\to e\gamma$ also required to
introduce the muon neutrino, prior to the $\nu_\mu$ discovery in 1962~\cite{PhysRevLett.9.36},
to cancel FCNC amplitudes~\cite{Feinberg:1958zzb}.

Radiative lepton decays $\ell_1\to\ell_2\gamma$ proceed via dimension-five left 
and right-handed radiative transition amplitudes. The branching fraction can 
be written in the form~\cite{Marciano:2008zz}
\beq
    \BR(\ell_1\to\ell_2\gamma) = \frac{3\alpha}{32\pi}\left(|A_L|^2 + |A_R|^2\right)
                                 \cdot \BR(\ell_1\to\ell_2\nu\nub)\,.
\eeq
For generic new physics at mass scale $\Lambda$ one can parametrise the left
and right-handed dipole amplitudes by $A_L=A_R=16\sqrt{2}\pi^2/G_F\Lambda^2$, 
where $G_F$ is the Fermi constant and $\Lambda$ the scale of the LFV interaction. 
The upper limit of $\BR(\mu\to e\gamma)<1.2\cdot10^{-11}$, obtained by the MEGA 
experiment at the Los Alamos Meson Physics Facility in 2001~\cite{Ahmed:2001eh},
thus translates into the stringent bound $\Lambda>340\:\tev$~\cite{Marciano:2008zz},
which is well beyond the LHC reach for direct detection. Decays involving virtual photons, 
such as $\ell_1\to\ell_2\overline\ell_2\ell_2$ and \mue conversion, have an additional 
rate suppression factor $\alpha_{\rm QED}$, but also probe different physics processes.

\begin{figure}[t]
\begin{center}
\includegraphics[width=0.8\columnwidth]{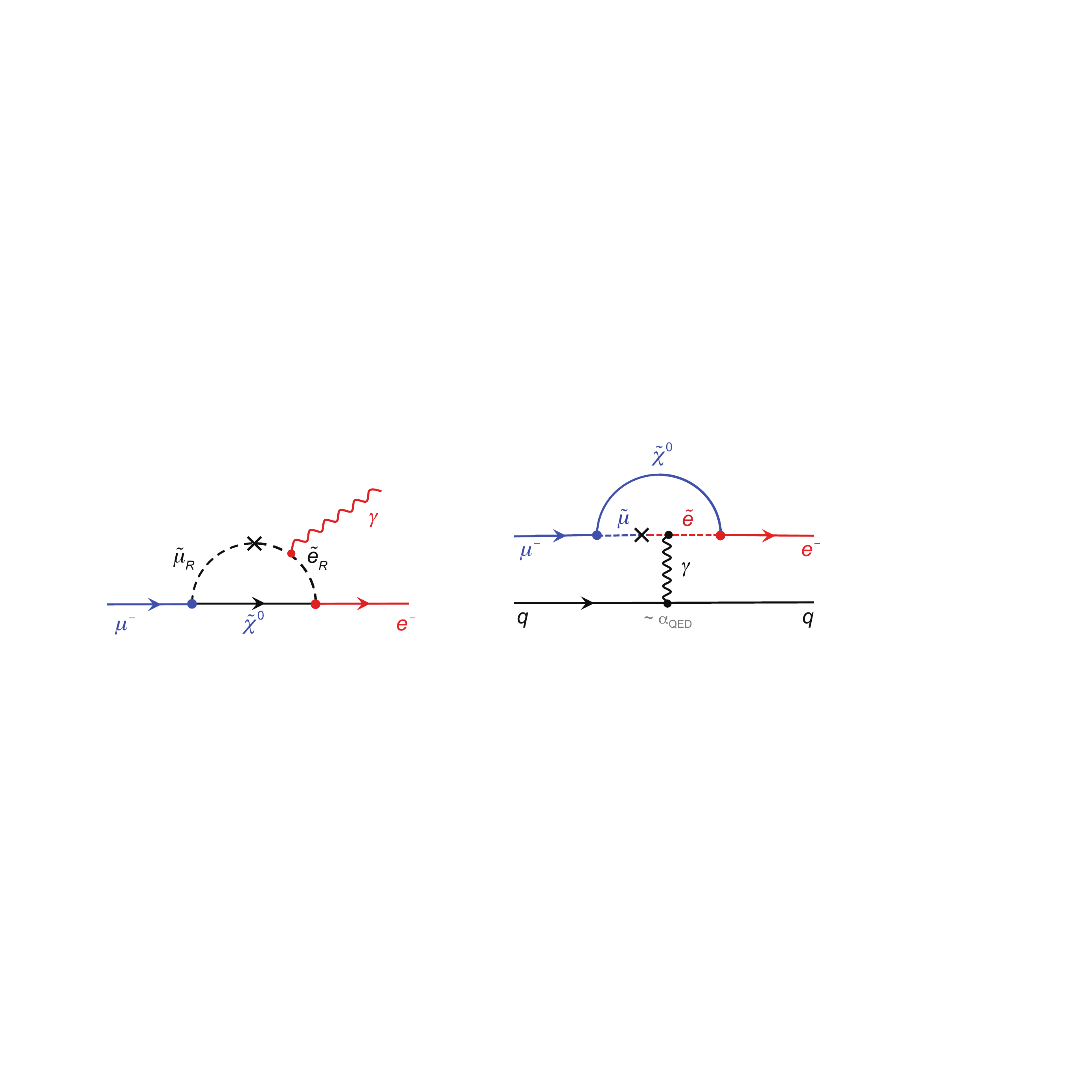}
\end{center}
\vspace{-0.3cm}
\caption{\label{fig:lfv-graphs}
  Possible supersymmetric contributions to the transition dipole diagrams
  mediating the LFV processes $\mu\to e\gamma$ (left) and $\mu N\to e N$ 
  conversion (right). }
\end{figure}

Figure~\ref{fig:lfv-graphs} depicts example graphs for $R$-parity conserving supersymmetric 
contributions to the charged LFV processes $\mu\to e\gamma$ (left) and $\mu N\to e N$ 
conversion (right). The predicted rates depend on the value of the slepton mass mixing 
parameter involved (\cf~\cite{Kuno:1999jp,Raidal:2008jk} and references therein).
Lepton flavour violation is also naturally present in $R$-parity violating models, 
where the strength of the effects is governed by the size of trilinear lepton number
violating couplings involving sleptons and leptons ($\lambda$), and squarks, leptons 
and quarks ($\lambda^\prime$) in the supersymmetric superpotential~\cite{Barbier:2004ez}. 

\subsection{A new limit on $\BR(\mu^+\to e^+\gamma)$ by the MEG experiment}

The MEG experiment~\cite{meg1,Adam:2009ci} uses the presently most powerful quasi-continuous 
muon beam produced at the PSI (Switzerland) $\pi$E5 beam line. Positive 29\:\mev surface 
muons hit with $3\cdot10^7\:$Hz rate a thin stopping target that is surrounded by 
the MEG detector. The muon decay rate measured by MEG effectively has no time 
structure, because the 2.2\:$\mu$s muon 
lifetime is long compared to the 50\:MHz radio-frequency structure of the 
proton cyclotron producing the muons. MEG consists of a positron spectrometer (drift
chamber) immersed in a gradient magnetic field that sweeps the produced positrons out 
of the interaction region, a time-of-flight counter, and a 900 litre liquid-xenon 
(LXe) scintillation detector outside of the magnet, measuring the photon incidence, time 
and energy. The solid-angle acceptance around the target is 10\%. 

The $\mu^+\to e^+\gamma$ signal events are characterised by back-to-back, in-time 
monoenergetic (52.8\:\mev) positron-photon pairs. Their measured energies, polar 
and zenith opening angles, and time difference are used to separate them 
from backgrounds, which are dominated by accidental coincidence of a positron 
from standard $\mu^+\to e^+\nu\nub$ decays and a photon from radiative 
$\mu^+\to e^+\gamma\nu\nub$ decays, bremsstrahlung or positron annihilation 
in flight. The reliance on a precise back-to-back signature invalidates
the use of negative muons, which would form muonium atoms in the target that
would smear out the two-body kinematics. The dominance by accidental 
background (that increases quadratically with the muon intensity), motivates 
the use of a less intense continuous beam rather than an intense pulsed beam. 
Ultimately, this background imposes a limitation on further progress in this 
channel with the current experimental techniques. 

\begin{figure}[t]
\begin{center}
\includegraphics[width=1.01\columnwidth]{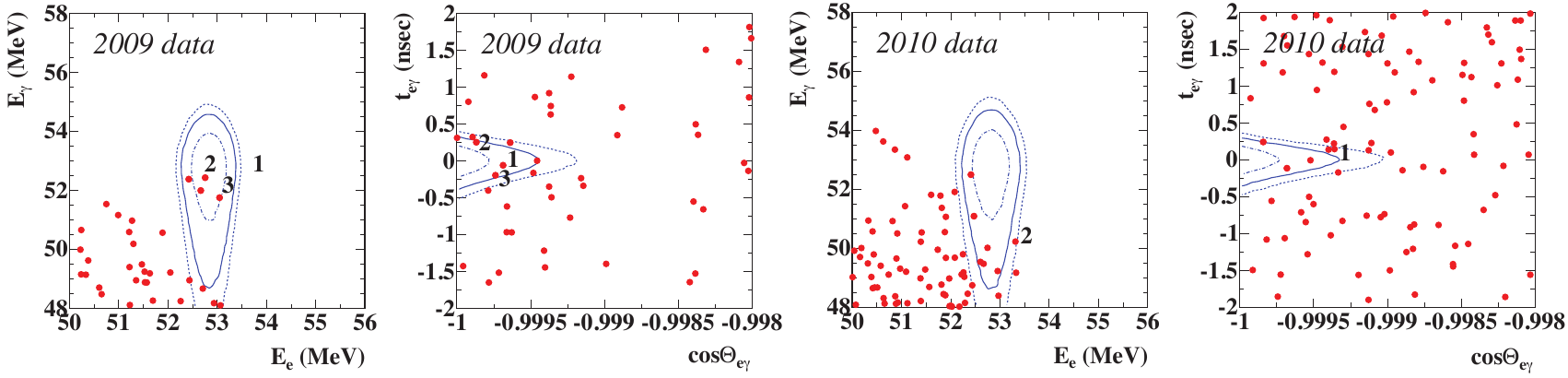}
\end{center}
\vspace{-0.3cm}
\caption{\label{fig:meg-likelihoods}
  MEG events (solid dots) in signal regions selected in 2009 (left two plots) and 2010 (right 
  plots). Shown are the measured photon versus positron energies and time difference versus 
  polar opening angle, respectively. The 1, 1.64 and 2$\sigma$ signal likelihood 
  contours are also drawn, and a few events with the highest signal likelihood are 
  numbered for each year. The figures are taken from~\cite{Adam:2011ch} (legend modified).}
\end{figure}

In 2010, the MEG Collaboration has released a preliminary result 
from the analysis of the 2009 data sample~\cite{Sawada:2010zz}, comprising 
$6\cdot10^{13}$ $\mu^+$ decays in the target. More than -- for the null hypothesis --
expected events in the signal region were found, leading to a larger than 
expected 90\% CL upper limit on $\BR(\mu^+\to e^+\gamma)$ of $1.5\cdot10^{−11}$,
compared $6.1\cdot10^{−12}$ expected. MEG has reanalysed the 2009 data in 2011, 
confirming that the excess of events is compatible with background-only at the 
8\% level, and also included the larger 2010 data set to a total $\mu^+$ yield 
of $1.8\cdot10^{14}$. The 2010 data benefited from an improved LXe waveform time 
resolution, but had slightly worse positron tracking resolution due to larger 
noise levels in the drift chamber~\cite{Adam:2011ch}. The two-dimensional energy, 
angular and timing distributions of the selected events in the signal regions for 
the 2009 (left) and 2010 data sets (right) are shown in Fig.~\ref{fig:meg-likelihoods}.
The 2010 data did not reproduce the 2009 excess of signal-like events. MEG quotes the
combined 90\% CL upper limit~\cite{Adam:2011ch}
\beq
   \BR(\mu^+\to e^+\gamma) < 2.4\cdot10^{-12}\,,
\eeq
compared to an expected limit of $1.6\cdot10^{-12}$.
This result is statistically limited. Systematic uncertainties are dominated by 
inaccuracies in the likelihood model and normalisation factors that should partly 
decrease with more statistics. The MEG experiment continues data taking in 2011 and 
2012 to explore the $\mu^+\to e^+\gamma$ decay down to the design sensitivity of several 
$10^{-13}$.

The authors of~\cite{Calibbi:2006nq} have explored the potential of the MEG experiment
(and others) in terms of supersymmetric GUT SO($10$) MSUGRA models, which embed the 
seesaw mechanism
and relate the neutrino Yukawa couplings to those of the up quarks making them naturally 
large~\cite{Masiero:2002jn} so that sizable LFV effects are introduced by the renormalisation 
group evolution. The ignorance of the Yukawa couplings between left and right-handed neutrinos 
is treated with the use of two extreme scenarios: minimal (CKM-like) and maximal (PMNS-like) 
mixing. Allowing MSUGRA parameter ranges of $m_0 < 5\:\tev$, $-3m_0 < A_0 < 3m_0$, 
sign$(\mu) = \pm$, it is found that the new MEG limit excludes the maximal mixing case for 
large values of \tanb and irrespective of the $m_{1/2}$ value (\cf Fig.~6 
in~\cite{Calibbi:2006nq}). 

\subsection{Muon to electron conversion}

The capture of a muon in an atom and its neutrinoless conversion to an electron in the 
recoil field of the nucleus is called \mue conversion (\cf right-hand graph in 
Fig.~\ref{fig:lfv-graphs} for a supersymmetric diagram of a \mue conversion process). 
In a \mue conversion experiment slow negatively charged muons are guided
via gradient magnetic fields to hit a stopping target where they are quickly captured 
by an atom (within $\sim$$10^{-10}\:$s) and cascade down to 1S orbitals. There, the $\mu^-$ 
either decays (with rate of $\sim$$5\cdot10^5\:$Hz), is weakly captured by the nucleus which 
undergoes $\beta$ decay (exceeding the muon decay rate for heavy atoms), or converts 
in an LFV process to an electron via virtual photon exchange with the nucleus. The \mue 
conversion gives rise to a single monoenergetic electron of 104.96\:\mev for Al atoms 
and 95.56\:\mev for Au, where the deviation from the muon mass is due to the muonic atom binding 
energy and the nucleus recoil energy. This simplicity and distinctive signature (very low 
background electron rate at 105\:\mev, no accidentals for single particle signature)
allows extremely high rates and thus a very sensitive measurement. 

The experimental observable is the conversion-to-capture ratio
\beq
\label{eq:raz}
    R_{\mu e}(A,Z) = \frac{\Gamma(\mu^- + N(A,Z) \to e^- + N(A,Z))}
                        {\Gamma(\mu^- + N(A,Z) \to \mbox{all $\mu^-$ captures})}\,,
\eeq
where $A$ and $Z$ are the atomic mass and charge number, respectively. The dependence 
of the ratio on the atom used can be exploited to distinguish new physics models 
after a discovery~\cite{Cirigliano:2009bz}. 
The denominator in Eq.~(\ref{eq:raz}) requires the knowledge of the total number of
muon captures that can be measured from standard processes. Owing to the virtual 
interaction of the lepton and nucleus systems, \mue conversion is sensitive
to classes of contact interaction models that do not contribute to 
$\mu\to e\gamma$~\cite{Raidal:2008jk}.
The conversion process might also exhibit enhanced sensitivity to supersymmetric Higgs 
exchange, which can be used to discriminate between models~\cite{Kitano:2003wn,Cirigliano:2009bz}.
On the other hand, the sensitivity to generic chirality flipping dipole LFV
amplitudes of new physics is reduced by factors 1/238, 1/342, 1/389 in branching 
ratios for Ti, Pb and Al targets, respectively~\cite{Czarnecki:1998iz,Kitano:2002mt,Marciano:2008zz}.
This requires a more precise measurement of $R_{\mu e}(A,Z)$ than that of $\BR(\mu\to e\gamma)$
for equal sensitivity, and hence $\mu^-$ rates of challenging $10^{11}$\:Hz.
Because no negative surface muon beams can be used, the beam will have a broader
momentum spectrum and is contaminated by in-time particle backgrounds, mainly pions.

The best current limit on \mue conversion has been obtained by the SINDRUM-II experiment
at PSI in 2006 using a gold target. They find $R_{\mu e}({\rm Au}) <7\cdot10^{-13}$ at 90\% 
CL~\cite{Bertl:2006up} (\cf Fig.~\ref{fig:lfv-history}). The extrapolated MEG sensitivity 
for $\mu^+\to e^+\gamma$ of $10^{-13}$ requires $R_{\mu e}(A,Z)\sim10^{-16}$. Such an accuracy 
is in reach of the proposed experiments Mu2e (FNAL, USA)~\cite{mu2e} and COMET 
(J-PARC, Japan)~\cite{comet}, which have both passed important approval steps. 
Their goal is to achieve a sensitivity down to $2$--$3\cdot10^{-17}$ by the year 2018, 
which requires to produce a total of $5\cdot10^{19}$ muons. 
The experimental approaches of Mu2e and COMET 
are similar, albeit with some specific differences. Mu2e uses pulsed beams
of 0.6\:MHz to eliminate prompt backgrounds (\eg, radiative pion capture), a pulse 
extinction of $10^{-10}$ after 0.7\:$\mu$s, and exploits the late muonic atom decays 
($\tau\sim1\:\mu$s) for the conversion measurement. A magnetic bottle traps pions 
produced by proton bombardment in a target, which decay into accepted muons. Gradient 
magnetic fields guide the muons through an S-shaped beam line from the production to the 
stopping target to increase the muon acceptance and momentum selection, and 
to evacuate loopers. Collimators reduce backgrounds from particles other than 
muons. For detection, the atoms pass through a tracking detector and are stopped
in a calorimeter. The decay electrons describe a helix trajectory in the gradient 
field of a solenoid where their momentum is precisely measured. The expected event 
yields for $R_{\mue} = 10^{-16}$ are $\sim$4 signal and 0.2 background events 
(dominated by radiative pion capture and decay-in-orbit muons). The multiple ring 
structure at FNAL allows to run Mu2e without interfering with the NOvA operation.
COMET uses a C-shaped solenoid for improved muon momentum selection prior to hitting
the stopping target, and also
a C-shaped detector section (before the tracker) to eliminate low-energetic 
decay-in-orbit muons. Fascinating long-term upgrade projects, using muon storage rings
that can improve the \mue conversion sensitivity by two orders of magnitude
would be offered by an experiment at the Project-X proton accelerator complex at 
FNAL~\cite{projectx} and by the PRISM/PRIME project~\cite{prism} at J-PARC (site not fixed). 

\subsection{Lepton flavour violation in $\tau$ decays}

\begin{figure}[t]
\begin{center}
\includegraphics[width=0.7\columnwidth]{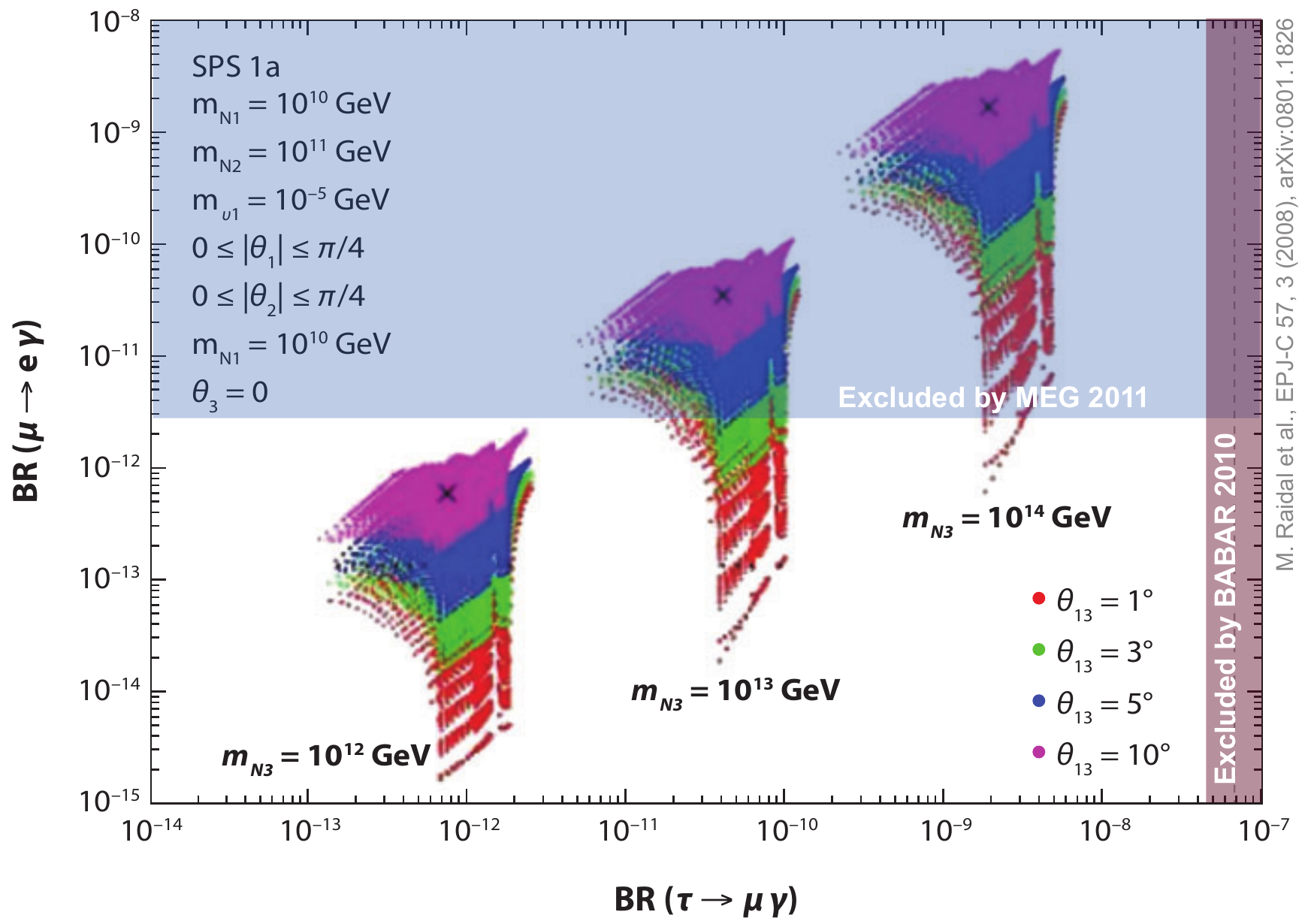}
\end{center}
\vspace{-0.3cm}
\caption{\label{fig:lfv-tau}
  Correlation between $\mu\to e\gamma$ and $\tau\to\mu\gamma$ branching fractions for
  various values of the third generation right-handed neutrino mass, $m_{N3}$,
  and the PMNS mixing angle, $\theta_{13}$, in a minimal supersymmetric seesaw model 
  and assuming the MSUGRA point SPS1a~\cite{Raidal:2008jk}. The shaded areas
  indicate the experimentally excluded regions. Figure taken from~\cite{Raidal:2008jk} 
  (modified). 
}
\end{figure}
The lower available fluxes of $\tau$ production in \ee accelerators renders the search 
for LFV in $\tau$ decays more challenging than for muons, which are conveniently 
produced from pion decays (\cf~\cite{Gonderinger:2010yn} for a discussion of $e$--$\tau$ 
conversion physics at an electron--ion collider). The best current sensitivity for 
$\tau\to\ell\gamma$ decays (charge averaged), based on a sample of almost one billion 
single $\tau$ decays, is obtained by the BABAR experiment setting 90\% CL upper limits of
$3.3\cdot10^{-8}$ (for $\ell=e$) and $4.4\cdot10^{-8}$ ($\mu$)~\cite{:2009tk}. The 
most stringent absolute upper limit of $1.8\cdot10^{-8}$  for LFV in $\tau$ decays has been 
obtained by the Belle 
experiment in the mode $\tau\to\mu\rho^0$~\cite{Miyazaki:2011xe}. See~\cite{Asner:2010qj}
for a full compilation of the results. Super {\em B}-factory projects in Italy~\cite{Bona:2007qt} 
and Japan~\cite{Abe:2010sj}, with anticipated integrated luminosities reaching up 
to 50\:ab$^{-1}$, can improve the $\tau\to\ell\gamma$ sensitivity down to a few 
$10^{-9}$~\cite{1742-6596-171-1-012079}.

What is the expected sensitivity to new physics of $\tau\to\ell\gamma$ compared to 
$\mu\to e\gamma$? This question can be addressed in specific new physics models 
by comparing the predicted branching fractions as a function of the model parameters. 
Such a study has been performed in~\cite{Raidal:2008jk} the results of which, for a 
minimal supersymmetric seesaw model and assuming the MSUGRA point SPS1a, is shown 
in Fig.~\ref{fig:lfv-tau}. The various regions correspond to different assumptions 
for the heavy neutrino mass and the $\theta_{13}$ mixing angle. If the recent T2K 
result~\cite{Abe:2011sj} on $\nu_\mu\to\nu_e$ appearance is confirmed (the significance 
of the current $\nu_e$ appearance excess is evaluated to be $2.5\sigma$), $\theta_{13}$ 
should be larger than $5^o$ hence effectively excluding the $m_{N3}=10^{14}\:\gev$ case. 
A large mixing angle leads to similar contributions to both LFV decays which renders 
the $\tau$ channel non-competitive. Similar results are found in~\cite{Hisano:2009ae}, 
where also a study using SU(5) without right-handed neutrinos is performed
exhibiting larger differences between the $\tau\to\ell\gamma$ and $\mu\to e\gamma$
branching fractions, although the predicted effects are smaller in general than for 
supersymmetric GUT seesaw models.

\section{Electric dipole moments}

Elementary particles are predicted to be non-spherical, distorted by an electric 
dipole moment (EDM). However, as for charged LFV, EDMs are predicted to be
undetectably tiny in the SM. Thus, any hint for a non-zero EDM would be a clean probe
for physics beyond the SM, and indeed many SM extensions predict EDMs that are detectable 
by current experiments.

Electric dipole moment are \CP-violating. The Hamiltonian of a system with magnetic 
moment, $\mu$, and EDM, $d$, immersed into magnetic and electric fields, $\vec B$ and $\vec E$,
is given by $H=-\mu\vec B\cdot \hat S -d\vec E\cdot \hat S$ (describing a magnetic
and electric Zeeman effect), where $\hat S$ is the unit
spin vector. Applying {\em P} and {\em T} transformations on $H$ we find that the product 
$\vec B\cdot \hat S$ transforms even under both {\em P} and {\em T}, while $\vec E\cdot \hat S$
transforms odd under these so that, assuming {\em CPT} invariance, a non-zero EDM is 
\CP non-conserving.

All experiments searching for EDMs follow the same basic principle. A particle with 
spin-1/2 immersed into magnetic and electric fields sees its spin vector precessing with
Larmor frequency $\omega_{\uparrow\uparrow}=2(\mu B + d E)/\hbar$, where the arrows indicate 
that both $B$ and $E$ fields have parallel orientation. Given the strong existing limits
on $d$, a small value of $d\sim10^{-26}\:e$cm in an electric field of 10\:kV/cm 
induces a tiny precession frequency of only 0.1\:$\mu$Hz. This frequency corresponds
to a magnetic field strength of 20 pico Gauss for a neutron~\cite{Lamoreaux:2009zz}. 
Given that already the magnetic field of the earth amounts to 0.2--0.7 Gauss, it is 
obviously impractical to measure the 
$E$-induced offset and thus $d$ directly. The effect of $B$ (if time independent)
can be cancelled by flipping the sign of the $E$ field and thus determining the 
frequency difference $\Delta\omega=\omega_{\uparrow\uparrow}-\omega_{\uparrow\downarrow}=4dE/\hbar$.
The sensitivity of the experiment thus depends on the control of the $B$ field 
variations between two $E$ flips, and on the $E$ field strength. As we will later see, 
the effective $E$ field seen by the dipole can be strongly amplified in 
paramagnetic atoms and molecules.

\begin{figure}[t]
\begin{center}
\includegraphics[width=0.8\columnwidth]{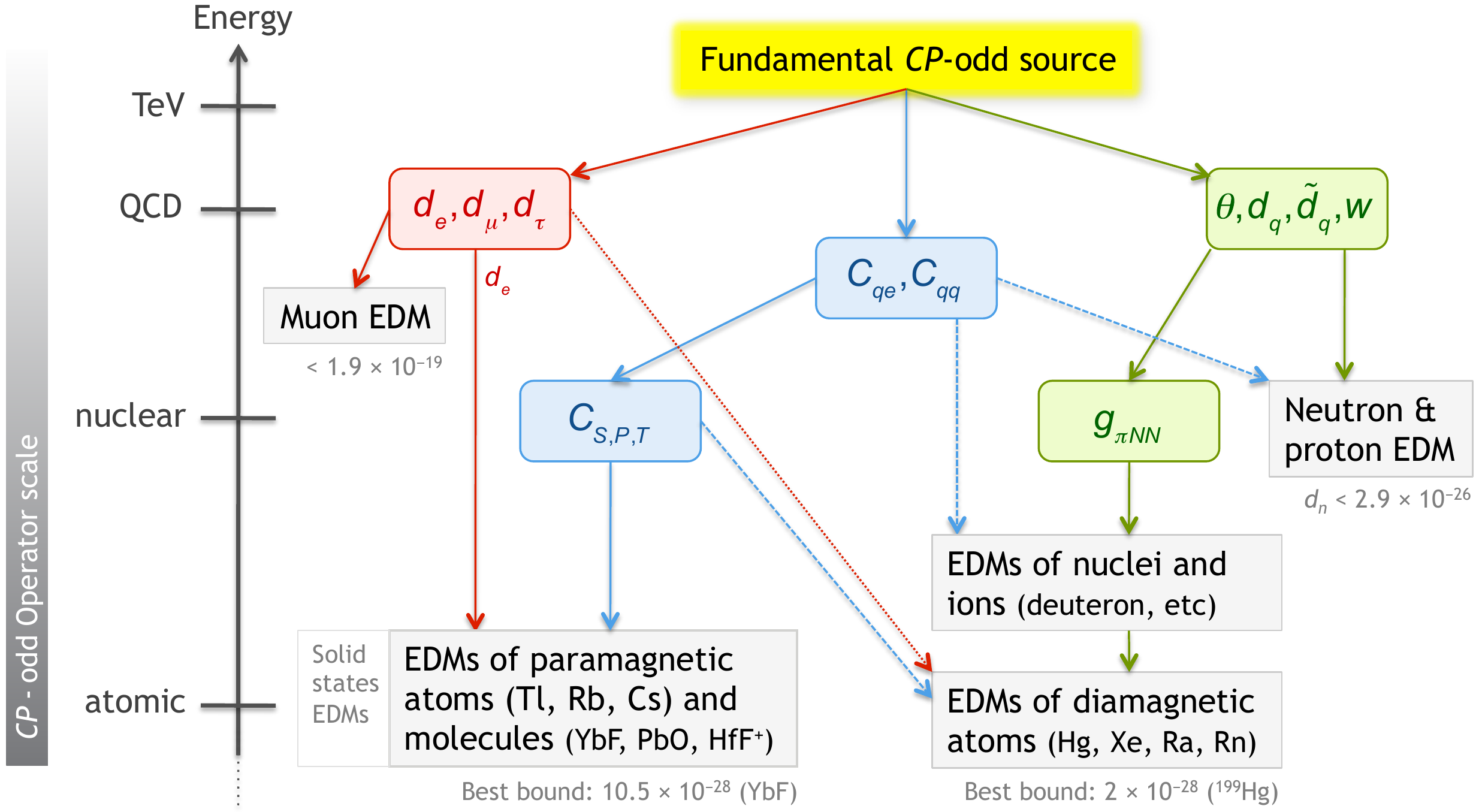}
\end{center}
\vspace{-0.2cm}
\caption{\label{fig:edm-plan}
   Hierarchy of scales between \CP-odd sources and generic classes of observable EDMs.
   For completeness one should also add solid state EDM effects to the atomic scale. 
   Figure inspired by~\cite{Pospelov:2005pr,Raidal:2008jk,KirchPanic}.
 }
\end{figure}
Figure~\ref{fig:edm-plan} depicts the hierarchy of scales between \CP-odd sources 
and generic classes of observable EDMs~\cite{Pospelov:2005pr}. On top, at \tev scale or 
beyond, the unknown source of \CP violation giving rise to lepton, light quark and colour 
EDMs at QCD scale, or a QCD $\theta$ term. The latter term is the only dimension-four operator 
and is thus not scale suppressed. This and the empirical smallness of $\theta$ is what 
is known as the {\em strong CP problem}. The EDMs formally belong to dimension-five operators, 
while most new physics models place them at dimension six (generating chirality flips) so that 
the EDMs are quadratically suppressed by the new physics scale. This makes them naturally 
small for high-scale new physics. The $w$ term in Fig.~\ref{fig:edm-plan} stands for 
a genuine dimension-six three-gluon operator. The $C_{ij}$ terms are coefficients of 
four-fermion operators, which are of dimension six or, in models with chirality flips, 
of dimension eight. These \CP-odd parameters are either directly observable, as is the 
case of the lepton EDMs, or induce \CP-odd operators at the nuclear scale. The most 
upfront observables at this scale are the neutron and proton EDMs, which directly 
arise from \CP-odd sources at the quark-gluon level. At the atomic scale, when dealing 
with nuclei, ions and atoms, also \CP-odd pion-nucleon interactions must be considered 
(denoted $g_{\pi NN}$ in Fig.~\ref{fig:edm-plan}). When EDMs are studied at the 
atomic scale, one distinguishes \CP-odd effects induced in paramagnetic and diamagnetic 
atoms or molecules. Paramagnetic atoms (named for their unpaired electrons, \eg thallium), 
and molecules (\eg ytterbium-fluorine), are mainly sensitive to the electron EDM and 
\CP-odd electron-nucleon couplings (denoted $C_{S,P,T}$ in Fig.~\ref{fig:edm-plan}). 
Diamagnetic atoms (\eg mercury) receive the same \CP-odd contributions as paramagnetic 
atoms but they are strongly suppressed due to smaller Schiff shielding violation, 
so that contributions from pion-nucleon scattering must also be considered.

\subsection{Neutron and proton EDMs}

\begin{figure}[t]
\begin{center}
\includegraphics[width=0.6\columnwidth]{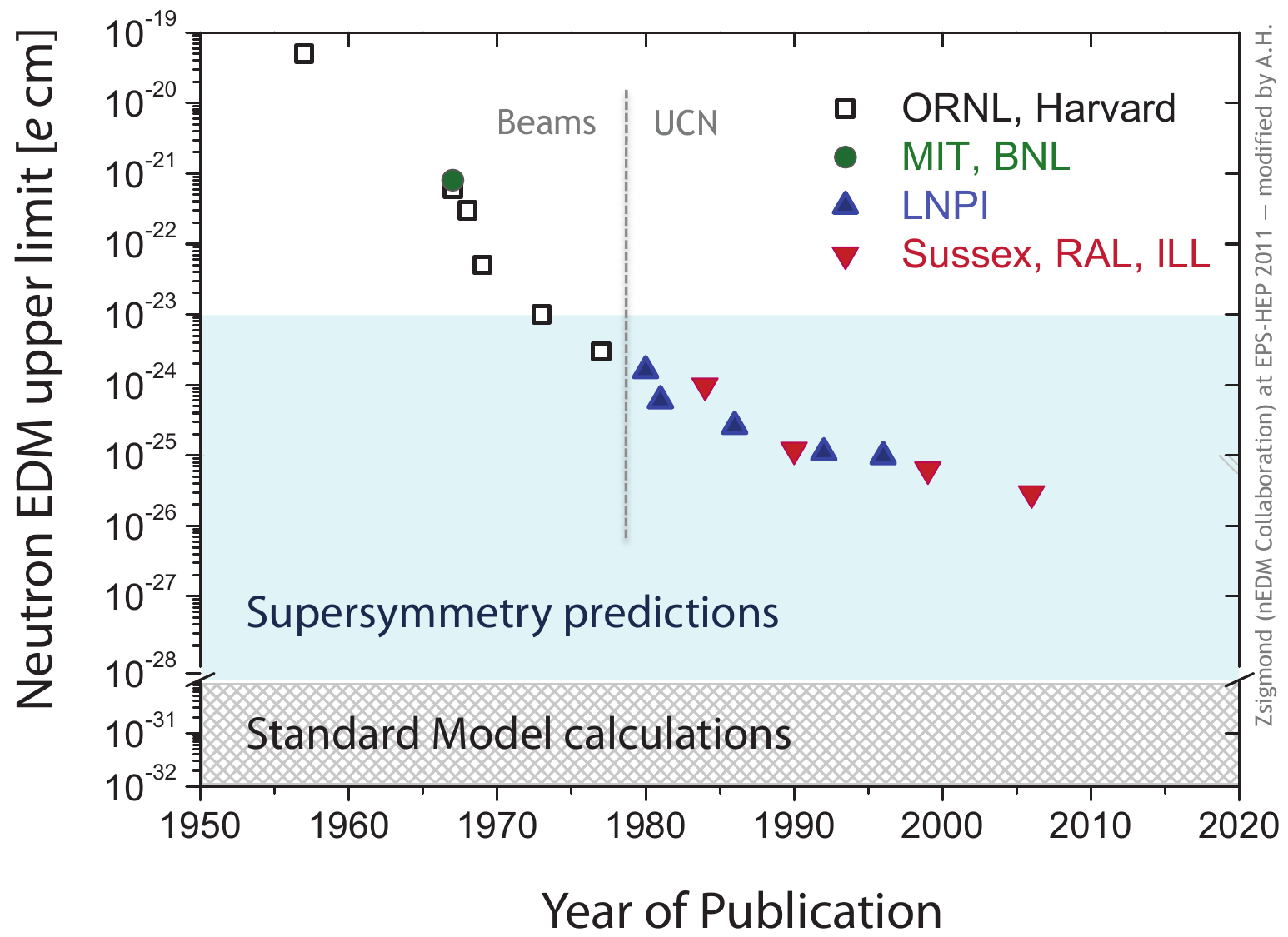}
\end{center}
\vspace{-0.2cm}
\caption{\label{fig:edm-neutron}
   History of neutron EDM measurements. The vertical line indicates the 
   switch to ultra-cold neutron facilities. Also shown are ball-park 
   ranges of generic supersymmetry and SM predictions. 
   Figure taken from~\cite{Zsigmond} (modified).
 }
\end{figure}
A chronological overview of the neutron EDM measurements is shown in 
Fig.~\ref{fig:edm-neutron}. The use of ultra-cold neutrons (UCN) 
provided a large sensitivity boost in this very active research field. Techniques 
such as in situ diamagnetic comagnetometers allowed to reduce systematic 
uncertainties due to magnetic field fluctuations. The currently best limit of 
$|d_n|<2.9\cdot10^{-26}\:e$cm at 90\% CL has been obtained by the Sussex-RAL-ILL 
experiment operating at the ILL-Grenoble, France~\cite{Baker:2006ts}. We can use it to
illustrate the strong \CP problem, $|\theta|\sim|d_n|\cdot2\cdot10^{16}<5\cdot10^{-10}$, and 
the supersymmetry problem, 
$|d_n|\sim10^{-23}\:e{\rm cm}\cdot(300\:\gev/M_{\rm SUSY})^2\cdot\sin\!\phi_{\rm SUSY}$,
where $M_{\rm SUSY}$ and $\phi_{\rm SUSY}$ are a generic supersymmetry mass scale and 
\CP-odd phase value, respectively (see, \eg,~\cite{Pospelov:2005pr}). Both parameters 
are subject to large suppression of unknown origin. 

Further improvement in the neutron EDM sensitivity requires to increase the UCN density, 
which can, for example, be achieved with the use of superfluid $^4$He used to moderate 
and store UCNs. Future experiments~\cite{KirchPanic} aim at sensitivities of (in units of 
$10^{-26}\:e$cm) $0.5$ and $0.05$ for the nEDM (by 2013) and n2EDM experiment at PSI (2016), 
respectively, $1$ for PNPI at ILL (2012), $0.3$ for CryoEDM at ILL (2016), $0.03$ for 
nEDM at SNS-ORNL (2020), $1$ for nEDM at RCNP (2014), $0.1$ and $0.01$ for nEDM at 
TRIUMF (2017 and $>$2020).

A fascinating perspective for measurements of the proton and/or light ion EDMs 
offer new proposals at BNL, USA~\cite{protonEDM-BNL} and 
J\"ulich, Germany~\cite{protonEDM-BNL}, with an anticipated sensitivity 
of $10^{-29}\:e$cm~\cite{Semertzidis}. The proposed experiments exploit a similar 
{\em magic-gamma} trick as used for the measurement of the muon anomalous magnetic 
moment in a homogeneous magnetic field, but here applied to a radial electric field 
that keeps the particles on orbit. A magic proton moment of 0.7\:\gev leads to 
aligned proton spin and momentum vector precession ($\omega_a=0$), thus allowing 
a precise EDM measurement via the monitoring of vertical spin precession versus
time. Assuming that the sensitivity to new physics is the same for protons and neutrons, 
these certainly are {\em must explore} experiments. 

\subsection{Electron EDM}

Over many years the electron EDM was dominated by the famous Berkeley measurement 
using thallium atoms that achieved in 2002 a 90\% CL upper limit of
$|d_e|<1.6\cdot10^{-27}\:e$cm~\cite{Regan:2002ta}. Again, this limit cuts into the 
bulk predictions of, \eg, supersymmetry, left-right symmetric and multi-Higgs models,
while the SM prediction lies even several orders of magnitude below that for the 
neutron EDM~\cite{eedm}. I will describe here a recent measurement performed at the IC-London, 
UK, using YbF molecules~\cite{Hudson:2011zz}.

Paramagnetic atoms and molecules respond to a screening theorem by 
Schiff~\cite{Schiff:1963zz} that implies a vanishing net EDM for a system built entirely 
from pointlike, nonrelativistic constituents that interact only electrostatically. 
Schiff's theorem is broken by magnetic and relativistic interactions, and by finite
size effects, \ie, a misalignment between the distribution of charge and EDM in 
the atom or molecule. Schiff violation is stronger for heavy atoms (growing as $Z^3$) 
and even stronger for polarisable molecules. It leads to an enhancement of the 
applied electric field, with factors of roughly $-585$ for the spherical thallium and 
$1.4$ million for the dipolar YbF. The YbF factor is given for an external $E$ field 
of $10$\:kV/cm, not rising linearly with the strength of the field due to saturation 
effects (see~\cite{Hudson:2002az} and references therein). Other advantages of YbF over 
Tl are smaller systematic uncertainties, in particular that due to motional magnetic 
fields. A shortcoming, however, is the much smaller production rate of YbF. 

\begin{figure}[t]
\begin{center}
\includegraphics[width=0.8\columnwidth]{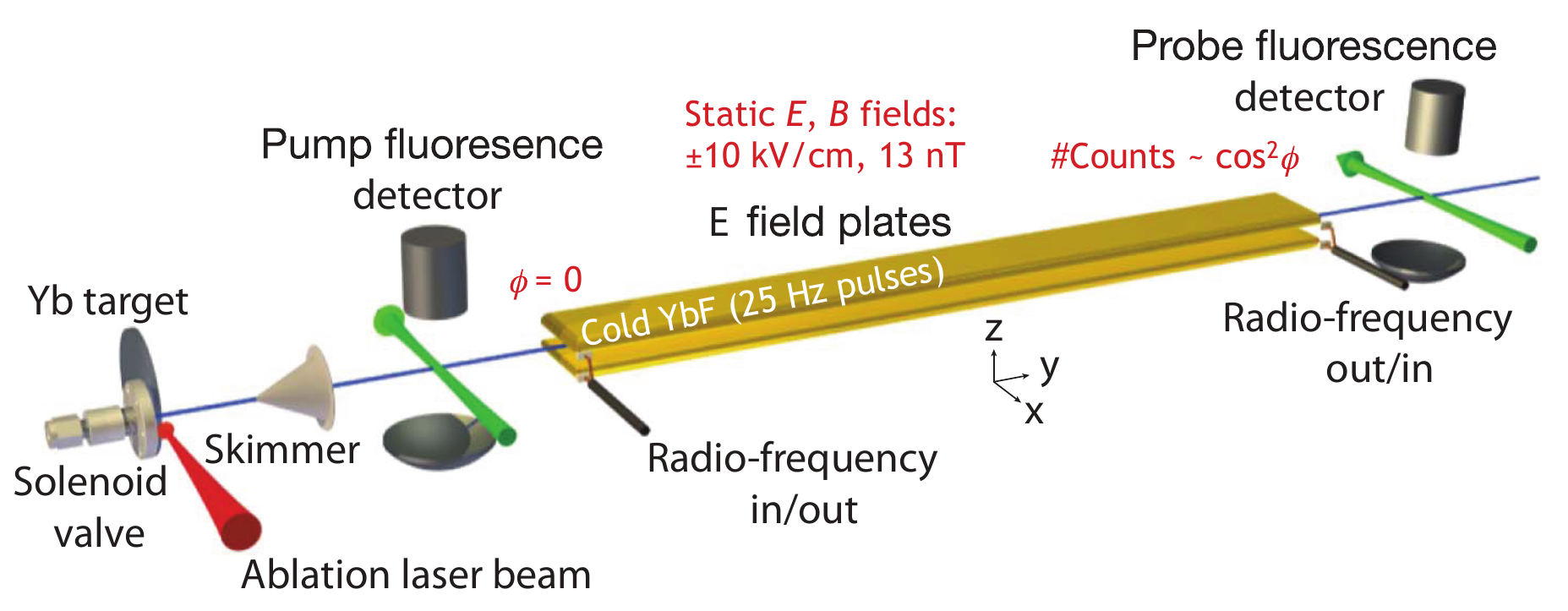}
\end{center}
\vspace{-0.2cm}
\caption{\label{fig:edm-ybf}
   Sketch of the ICL YbF electron EDM experiment~\cite{Hudson:2011zz} (modified).
   Pulsed YbF molecules travel from the left to the right in-between the electric 
   field plates.
 }
\end{figure}
The ICL YbF electron EDM experiment looks for a spin interferometer phase shift of
the $F = 0,1$ YbF hyperfine levels when the electric field is reversed. The principle 
of the table-top apparatus is depicted in Fig.~\ref{fig:edm-ybf} (taken 
from~\cite{Hudson:2011zz}). Laser ablated YbF molecules populating both 
hyperfine levels are depleted from $F=1$ states via optical pumping using the 
$552$\:nm optical transmission line of YbF. The remaining molecules in the $F=0$ 
ground state enter a pair of electric field plates that are magnetically shielded. 
A tuned $170\:$MHz radio-frequency (RF) pulse transfers the molecules from the 
ground state into the $F=1$ hyperfine level, where they form a coherent 
$(|F=1,m_F=1\rangle + |1,-1\rangle)/\sqrt{2}$ state. While travelling
some time $T$, the magnetic and, if non-zero EDM, electric fields induce a phase 
shift~\cite{Hudson:2011zz} $(e^{i\phi}|1,1\rangle + e^{-i\phi}|1,-1\rangle)/\sqrt{2}$, 
where $\phi=(\mu_B B-d_e E_{\rm eff})\cdot T/\hbar$. Another 
RF pulse projects the phase-shifted molecules back to $F=0$ with state function
$(e^{i\phi}+e^{-i\phi})|0,0\rangle/2$. A fluorescence pump detector counts the $F=0$ 
population that has a rate proportional to ${\rm cos}^2\phi$. One can scan the 
counting rate and thus $\phi$ by modulating the external magnetic field, and generate 
a tiny observable phase shift $\Delta\phi=2d_e E_{\rm eff}T/\hbar$ by reversing the 
electric field. In practise, the $B$ and $E$ fields as well as several other 
parameters of the apparatus are switched in a random sequence between each 
beam pulse to track and minimise systematic correlations.

A total of $25$ million YbF beam pulses, distributed throughout six thousand 
data blocks taken in 2010, give eight statistically independent 
measurements corresponding to manual reversal states of the apparatus. Combined 
they yield $d_e=(2.4 \pm 5.7_{\rm stat} \pm 1.5_{\rm syst})\cdot10^{-28}\:e$cm, and 
a 90\% CL upper limit of $|d_e|<10.5\cdot10^{-28}\:e$cm. The systematic error 
is currently dominated by the uncertainty in the uniformity of $E$ between sign
flips~\cite{Hudson:2011zz}.
This pioneering measurement has great potential in spite of the, yet, relatively little 
improvement over the 2002 thallium result. The present result is statistically limited, 
and systematic errors are expected to be reducible to a level smaller than $10^{-29}\:e$cm.
The ICL group aims at a factor of ten sensitivity improvement within a few years, 
with the final goal of reaching a factor of hundred improvement. Several other 
EDM experiments, based on electron spin precession in atoms, molecules, molecular 
ions or solids are under development.

\section{Tau-lepton physics}

Last year celebrated the 20$^{\rm th}$ anniversary of the Tau Workshop series which 
started at LAL-Orsay in 1990. During that period there was magnificent progress in 
tau-lepton physics through the LEP, CLEO, {\em B}-factory, BES, VEPP-2M, and 
neutrino experiments. Early Tau Worskhops concentrated on the consolidation of the 
tau as a standard lepton without invisible decays and with universal couplings. 
Increased data samples as well as better understanding and methods allowed the 
experiments to study electroweak and QCD physics with tau leptons leading to 
precision measurements of fundamental SM parameters such as ${\rm sin}^2\theta_W$, 
\as, and $|V_{us}|$. 

Whereas the four LEP experiments (1989--2000, CERN, Switzerland) collected each a 
data sample of roughly 165 thousand $Z\to\tau\tau$ events, CLEO (1979--2008,
Cornell, USA) disposed already of about 3.6 million tau pairs, albeit with less 
efficient and pure selection than at LEP. The {\em B}-factory experiments 
BABAR at SLAC, USA and Belle at KEK-B, Japan accumulated breathtaking 500 
million and 900 million tau pairs, respectively. Consequently, these experiments 
concentrate on the measurement of rare modes and searches for new physics:
lepton flavour violation in tau decays, charged weak current universality tests, 
rare branching fractions, second class currents (isospin violation), and 
\CP violation. Phenomenological work on the determination of \as and $|V_{us}|$ 
from tau branching fractions and spectral functions are also actively 
pursued.

\subsection{Tau branching fractions and tests of universality}

\begin{figure}[t]
\begin{center}
\includegraphics[width=0.6\columnwidth]{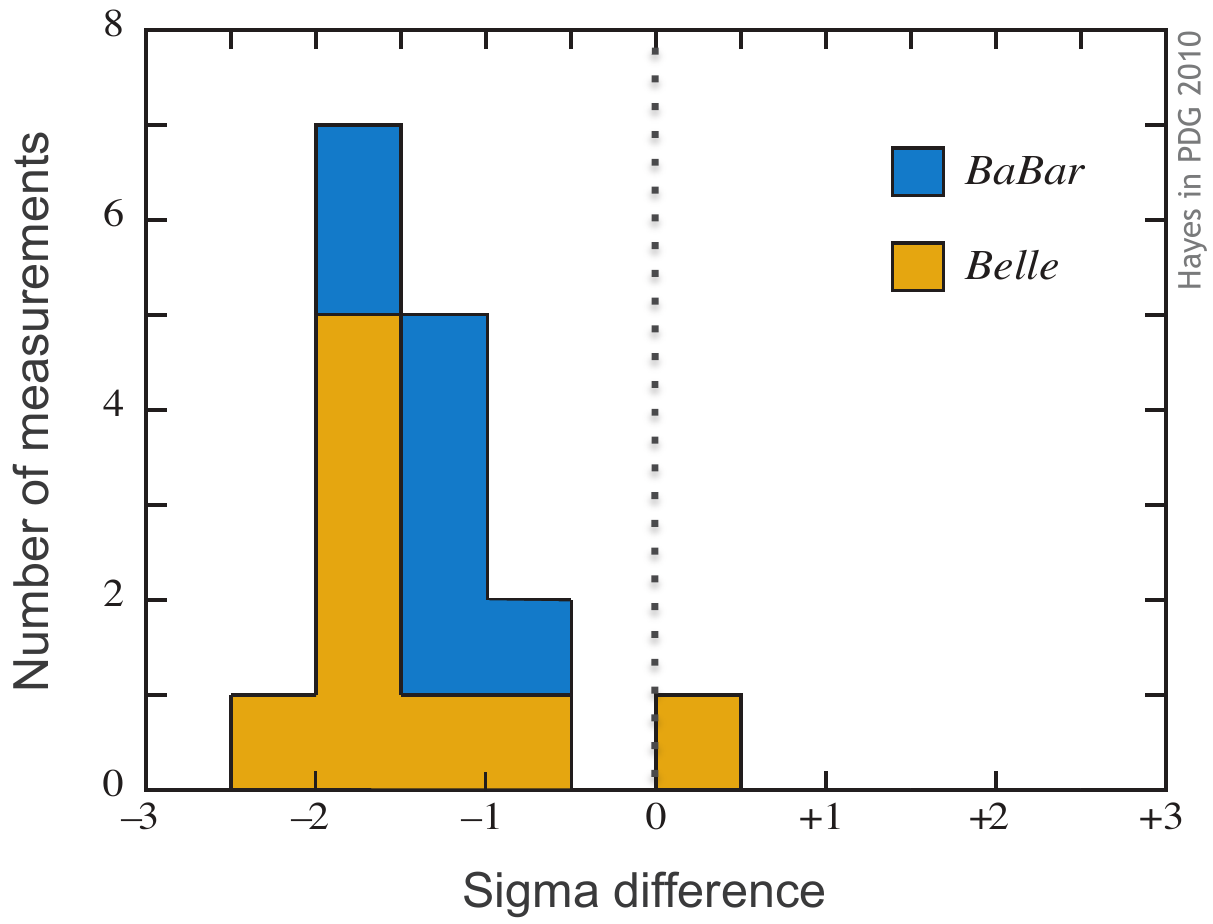}
\end{center}
\vspace{-0.2cm}
\caption{\label{fig:tau-brs}
   Pull difference between the 16 {\em B}-factory measurements of tau branching fractions 
   and their non-{\em B}-factory measurements. Belle and BABAR have each published 
   8 measurements. Figure from~\cite{Nakamura:2010zzi} (modified).
 }
\end{figure}

The most recent summary of the tau branching fractions is available from the 
Heavy Flavour Averaging Group (HFAG)~\cite{Asner:2010qj} (see tau section).
The dominant leptonic and hadronic tau branching fractions have been measured 
to great precision at LEP while running at the $Z$ resonance. The ALEPH experiment 
in addition performed a global analysis of all tau hadronic modes imposing 
unitarity~\cite{Schael:2005am}. The {\em B}-factory experiments also contributed
with several new measurements, in particular improving the modes with kaons owing
to their superior particle identification capabilities.  The PDG 
group~\cite{Nakamura:2010zzi} noticed a puzzling tendency when comparing the total 
of 16 {\em B}-factory measurements (upper limits and exotic searches not included) 
with their counterparts from LEP: all but one {\em B}-factory results lie below those 
from LEP (\cf Fig.~\ref{fig:tau-brs}). A feature that requires further scrutiny. 

The measurements of the branching fractions involving leptons allow one to test 
the three-generation universality of the charged weak current.
The HFAG finds for the weak coupling ratio $|g_\mu/g_e|$, obtained from 
the ratio of the $\tau\to\mu\nub\nu$ to $\tau\to e\nub\nu$ world average branching 
fractions (corrected for mass-dependent effects), $1.0018\pm0.0014$~\cite{Asner:2010qj},
which is a factor of seven more precise than the measurement of the corresponding
on-shell $W$ leptonic branching fraction ratio~\cite{Nakamura:2010zzi}. 
It is of similar precision as the universality tests performed in neutral current
reactions at LEP, giving $0.28\%$ from a comparison of the $Z\to\ell\ell$ branching fractions,
and $0.13\%$ from comparing the effective weak axial coupling $g_{A,e/\mu}$~\cite{:2005ema}.
Universality of the ratios $|g_\tau/g_e|$ and $|g_\tau/g_\mu|$ can be tested by comparing 
the measured tau leptonic branching fractions with the predicted ones using the 
tau and muon lifetimes and correcting for mass and radiative effects. The HFAG 
computes the world average ratios $1.0027\pm0.0021$ and $1.0008\pm0.0021$, respectively. 
Finally, the ratio $|g_\tau/g_\mu|$ can also be precisely determined with the ratio of 
the $\tau\to h\nu$ to $h\to\mu\nu$ ($h=\pi,K$) branching fractions, corrected for mass 
and radiative effects, which gives $0.9949\pm0.0029$. None of these tests reveals a 
significant deviation from universality. 

\subsection{Second class currents}

In tau decays the charged weak current is classified according to its $G$-parity,
where $J^{PG}=0^{--},1^{+-}$ currents ($\pi$, $a_1$, \dots) produce an odd number 
of pions in the final state, and $J^{PG}=1^{-+}$ currents ($\rho$, \dots) produce an 
even number of pions. These are denoted first class currents. 
{\em Second class currents}~\cite{Weinberg:1958ut} (SCC) have opposite spin-parity 
$J^{PG} = 0^{+-}, 0^{-+}, 1^{++},$ or $1^{--}$. They violate isospin symmetry and thus vanish
in the strict isospin limit. Their branching fractions are expected to be proportional 
to the $u,d$ quark mass difference-squared. Examples for SCC are the decays
$\tau\to\eta\pi\nu$ and $\tau\to\omega\pi\nu$, which may be dominantly mediated
by the $a_0(980)$ and $b_1(1235)$ resonances, respectively. Estimates predict 
$\BR(\tau\to\eta\pi\nu)=(1.3\pm0.2)\cdot10^{-5}$ (see~\cite{Pich:1987qq,Nussinov:2008gx} 
and references therein). No SCC mode has been observed to date. The best current 
experimental limit of $9.9\cdot10^{-5}$ at $90\%$ CL has been obtained by BABAR 
using the full available statistics~\cite{delAmoSanchez:2010pc}. Since this search
is not free of background a much larger data sample is required to be able to 
discover this mode if the branching fraction is as expected. 

\subsection{CP violation in tau decays}

\CP violation in tau decays with strangeness occurs in the SM via 
mixing-induced \CP violation in the neutral kaon system. The asymmetry 
\beq
     A_Q = \frac{\Gamma(\tau^+\to\pi^+\KS\nub_\tau)-\Gamma(\tau^-\to\pi^-\KS\nu_\tau)}
                {\Gamma(\tau^+\to\pi^+\KS\nub_\tau)+\Gamma(\tau^-\to\pi^-\KS\nu_\tau)}\,,
\eeq
is predicted to be $A_Q^{\rm SM}\simeq 2{\rm Re}(\e_K)=(0.33\pm0.01)\%$~\cite{Bigi:2005ts}.
Reference~\cite{Grossman:2011zk}, which appeared after the conference, points out that 
this prediction~\cite{Bigi:2005ts} for the \KS-only terms contains a sign mistake and that 
only when taking into account \KS--\KL interference the positive sign is approximately 
recovered. Deviations from $A_Q^{\rm SM}$ should have new physics origin as is possible, \eg, 
in multi-Higgs models~\cite{Kuhn:1996dv}. Results consistent with the SM were previously
found in $D^\pm\to\KS\pi^\pm$~\cite{delAmoSanchez:2011zza} and in an angular analysis of 
$\tau^\pm\to\pi^\pm\KS\nu$~\cite{Bischofberger:2011pw}.

For the conference, BABAR released a new result for $A_Q$~\cite{BABAR:2011aa}
based on the full available data set ($476\:{\rm fb}^{-1}$). The analysis uses 
electron and muon tagged events (recoil side). The raw asymmetry is corrected 
for different nuclear interaction cross sections and for feed-through from 
$\KS K^\pm(\ge\pi^0)$ and $\Kz\Kzb\pi^\pm)$ decays, where $A_Q^{\rm SM}$ is assumed
for the latter correction. BABAR finds $A_Q=(-0.45\pm0.24\pm0.11)\%$, which 
deviates by approximately $3\sigma$ from $A_Q^{\rm SM}$. However, as argued 
in~\cite{Grossman:2011zk}, the precise prediction of $A_Q$ as measured by BABAR 
must account for the decay time interval over which it is measured, which depends 
on the experimental conditions and analysis requirements applied. 

\subsection{Tau hadronic spectral functions and QCD}

QCD studies using the tau hadronic width and spectral moments computed from 
the tau hadronic spectral functions have traditionally been the most active field 
of phenomenological studies involving tau physics (see, \eg,~\cite{Davier:2005xq} 
for a review). In particular the measurements
of the complete vector and axial-vector tau hadronic spectral functions by 
ALEPH~\cite{Barate:1997hv,Barate:1998uf,Schael:2005am} 
and OPAL~\cite{Ackerstaff:1998yj} triggered widespread interest. 
Most prominently, tau decays allow an accurate determination of \as from the 
comparison of the non-strange tau hadronic width, 
$R_{\tau,V+A}=3.4771 \pm 0.0084$~\cite{Asner:2010qj}, 
precisely obtained using unitarity and universality from the measurements of 
the tau leptonic branching fraction and the tau lifetime, and subtracting the 
strangeness contribution, with the SM prediction from essentially perturbative QCD. 
Small nonperturbative contributions are fitted from data using the spectral 
moments. Quark-hadron duality violations are usually assumed to be negligible 
due to the suppression of the spectral moments at the tau mass scale (see the critical
discussion in~\cite{Boito:2011qt,Boito:2011pr}). The perturbative QCD prediction of $R_{\tau,V+A}$ 
benefits, as the electroweak fit to the $Z$ hadronic width, from an NNNLO calculation 
of the massless perturbative Adler function~\cite{Baikov:2008jh}. Unfortunately, 
an ambiguity in the perturbative treatment does currently not allow to fully 
exploit the available 
precision~\cite{Davier:2008sk,Beneke:2008ad,Menke:2009vg,Caprini:2011ya} in 
tau decays.

\begin{figure}[t]
\begin{center}
\includegraphics[width=0.7\columnwidth]{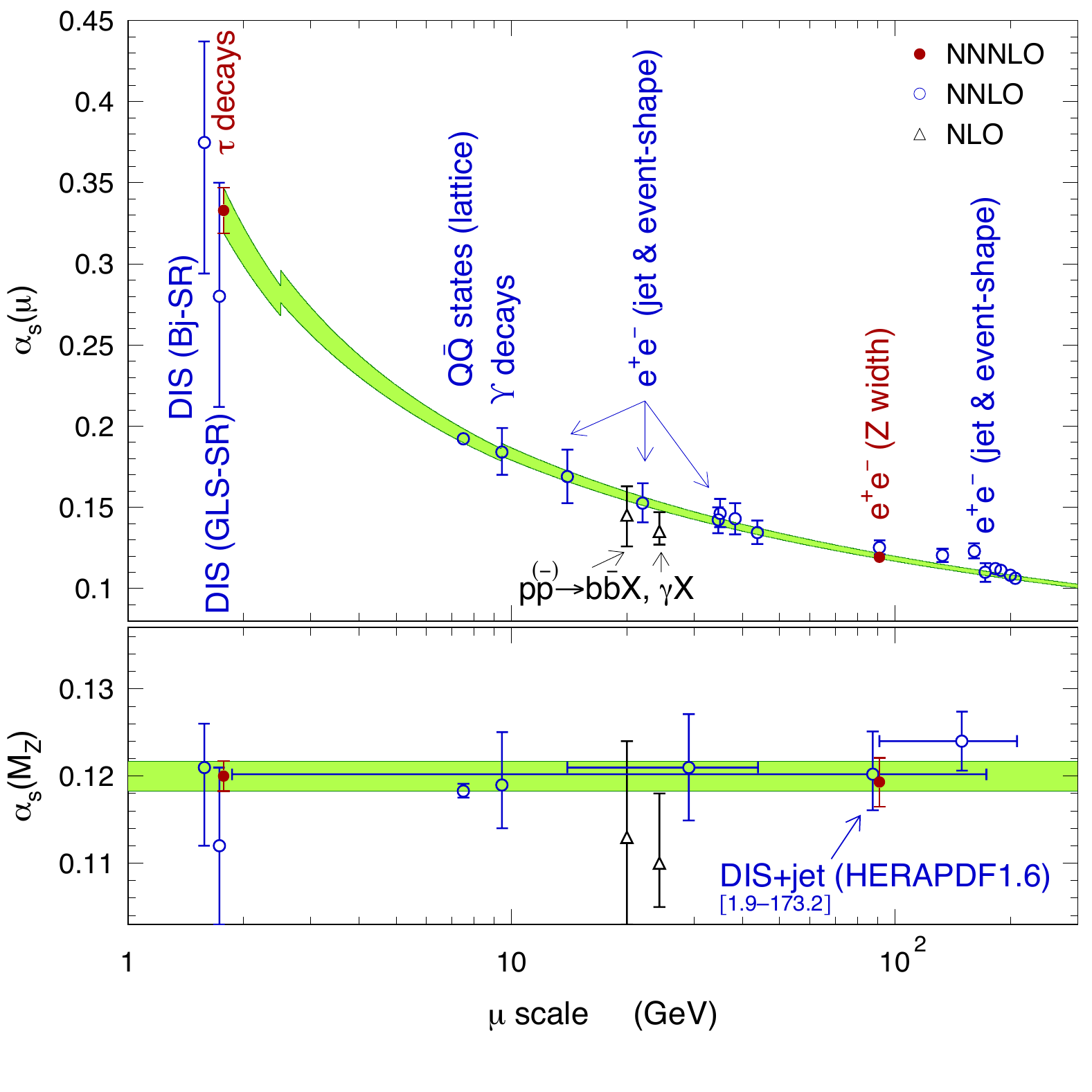}
\end{center}
\vspace{-0.7cm}
\caption{\label{fig:evolution_lp11}
   Top plot: evolution of \asTau given in Eq.~(\ref{eq:as}) 
   to higher scales $\mu$ using the four-loop RGE 
   and the three-loop matching conditions applied at twice the heavy quark-pair 
   thresholds (hence the discontinuities). The evolution is compared with independent 
   measurements (taken from the compilation~\cite{Bethke:2009jm}), covering $\mu$ 
   scales that vary over more than two orders magnitude. Bottom: the corresponding 
   $\as$ values evolved to $\mZ$. The shaded band displays the tau decay 
   result within errors.   
   Figure courtesy of Z. Zhang.
 }
\end{figure}

From the combined fit of \as and the nonperturbative terms to $R_{\tau,V+A}$ and 
several spectral moments, and after the renormalisation group evolution from the 
tau to the $Z$ mass, one finds
\beq
\label{eq:as}
   \as(M_Z) = 0.1200 \pm 0.0005_{\rm exp} \pm 0.0008_{\rm theo}
                     \pm 0.0013_{\rm CIPT/FOPT} \pm 0.0005_{\rm evol}\,,
\eeq
where the first error is experimental, the second theoretical, the third accounts
for half of the difference between the two perturbative treatments (denoted as 
contour-improved and fixed-order perturbation theory, of which the average result 
is used as central value for $\as(M_Z)$), and the last error accounts for the 
evolution uncertainty (four-loop evolution~\cite{vanRitbergen:1997va} with 
three-loop quark-flavour matching~\cite{Chetyrkin:1997sg,Chetyrkin:1997un,Rodrigo:1997zd}).
The nonperturbative contributions in~(\ref{eq:as}) are taken from~\cite{Davier:2008sk}.
Equation~(\ref{eq:as}) is in excellent agreement with the NNNLO result from
the electroweak fit to (mainly) the $Z$ hadronic width 
$\as(M_Z)=0.1193\pm0.0028$~\cite{Baak:2011ze}, providing a precise test of 
the asymptotic freedom property of QCD. The evolution path of \asTau to higher scales
is shown in the upper plot of Fig.~\ref{fig:evolution_lp11}. The evolution is compared 
in this plot with other \as determinations compiled in~\cite{Bethke:2009jm}.

\subsection{Determination of $\,|V_{us}|$}

The amount of net strangeness production in tau decays is directly proportional 
to the CKM element $|V_{us}|^2$. Accurate extractions of this element have been 
performed with the use of three different methods: 
$(i)$ comparing the measured branching fraction of the decay $\tau^-\to K^-\nu$ 
with its SM prediction, which depends on the kaon decay constant taken from Lattice QCD
and on radiative corrections, 
$(ii)$ comparing the measured ratio of the $\tau^-\to K^-\nu$ and $\tau^-\to \pi^-\nu$ 
branching fractions to the SM prediction, and 
$(iii)$ comparing the inclusive tau strange hadronic width with its SM 
prediction. The latter two methods also depend on $|V_{ud}|^2$. 

\begin{figure}[t]
\begin{center}
\includegraphics[width=1.00\columnwidth]{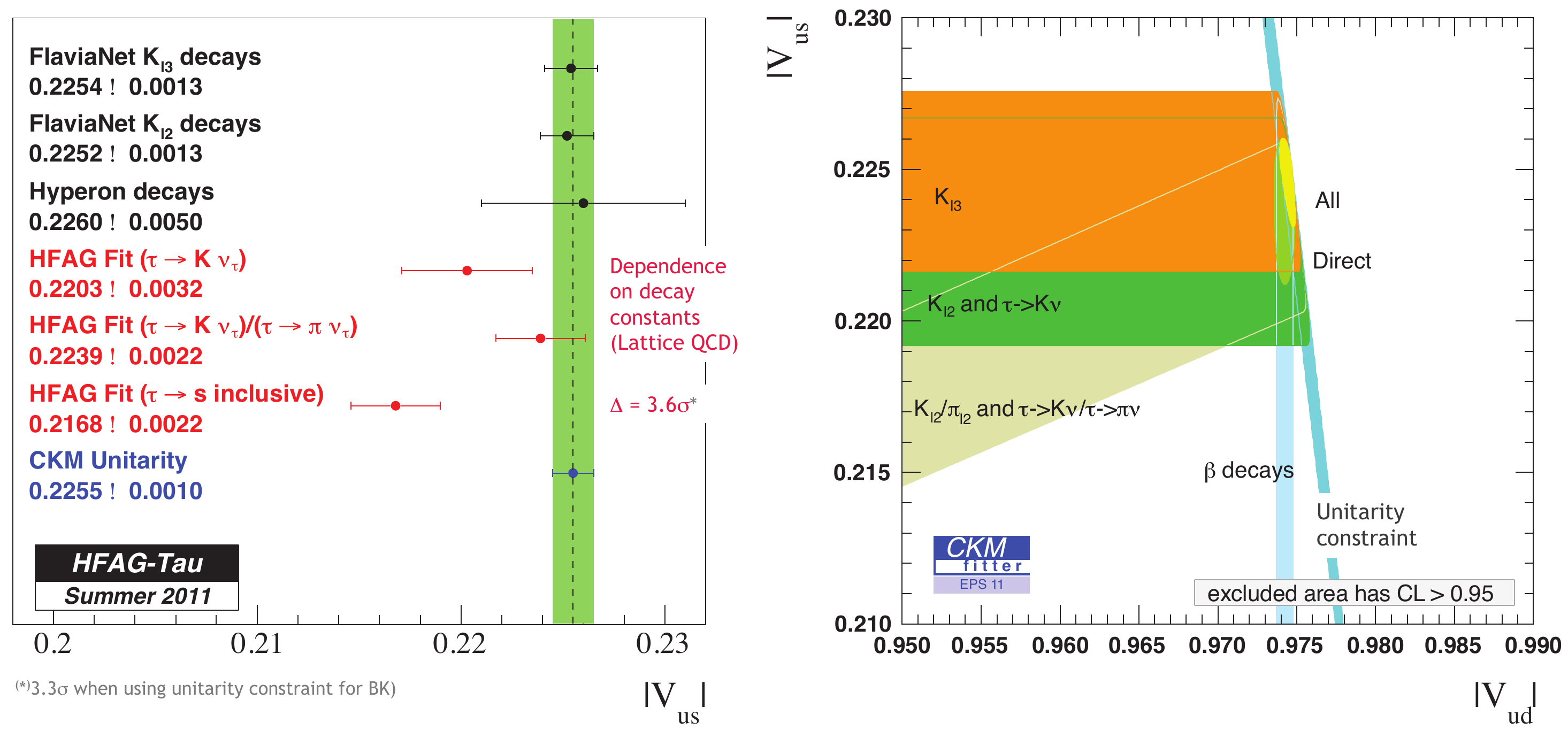}
\end{center}
\vspace{-0.2cm}
\caption{\label{fig:tau-vus}
   Left: determination of $|V_{us}|$~\cite{Asner:2010qj}. 
   The results from tau decays are printed in red. Right: $|V_{us}|$ versus $|V_{ud}|$.
   Shown are recent experimental determinations of these quantities and the constraint 
   from unitarity in the first row of the CKM matrix~\cite{ckmfitter-vus}. 
 }
\end{figure}

The results for $|V_{us}|$ obtained from these methods are shown in the left panel 
of Fig.~\ref{fig:tau-vus}, and are compared to other evaluations and to the CKM 
fit result assuming unitarity~\cite{Asner:2010qj}. The 
agreement between the tau inclusive result and the CKM fit is marginal. The 
inclusive method~\cite{Gamiz:2004ar} uses the observable 
$|V_{us}|^2=R_{\tau,S=1}/(R_{\tau,S=0}/|V_{ud}|^2-\delta R_{\tau}^{\rm SM}(\as,m_s))$~\cite{Barate:1999hj},
where the SM prediction of the mass-dependent term, $\delta R_{\tau}^{\rm SM}(\as,m_s)$,
exhibits slow convergence in the perturbative series~\cite{Chetyrkin:1993hi,Maltman:1998qz}, 
which leads to a hard-to-quantify theoretical uncertainty. The inclusive method 
may also suffer from potentially unmeasured modes, which are not included in the 
systematic uncertainties. Finally, the trend to smaller branching fractions from 
the {\em B}-factory experiments has a non-negligible impact on this quantity. 
Without the {\em B}-factory data,
$|V_{us}|$ from the inclusive determination would increase to approximately $0.2213$. 
A precise measurement of the full strange spectral function by the {\em B}-factory
experiments should help to test the inclusive method. 

The right panel of Fig.~\ref{fig:tau-vus} shows the experimental results on $|V_{us}|$ 
versus $|V_{ud}|$ compared to the CKM fit constraint~\cite{ckmfitter-vus}. Good agreement 
with unitarity of the first row is observed.

\section{Anomalous magnetic moment of the muon}

The Dirac equation predicts for elementary leptons, $\ell$, a magnetic moment,
$\vec{M}=(g_\ell e/2m_\ell)\cdot\vec{S}$, with gyromagnetic ratio $g_\ell=2$~\cite{review}.
Quantum fluctuations lead to a small calculable deviation from $g_\ell=2$, parametrised 
by the anomalous magnetic moment $a_\ell \equiv (g_\ell-2)/2$.
That quantity can be accurately measured and, within the SM, precisely predicted. 
Hence, comparison of experiment and theory tests the SM at its quantum loop level. 
A deviation in $a^{\rm exp}_\ell$ from the SM expectation would signal effects of new 
physics, with current sensitivity reaching up to mass scales of 
${\cal O} ({\rm TeV})$ for $\ell=\mu$~\cite{Czarnecki:2001pv,Davier:2004gb}. Considering 
the expected quadratic dependence of new physics contributions on the lepton 
mass~\cite{Czarnecki:2001pv}, and the current experimental uncertainties on 
$a^{\rm exp}_\ell$, the muon $g-2$ is roughly 50 times more sensitive to new physics
than the electron one,\footnote
{
   In spite of the breathtaking accuracy of the most recent electron $g-2$ 
   measurement by the Harvard group~\cite{Hanneke:2008tm},
   giving $a^{\rm exp}_e=(11\,596\,521\,807.3 \pm 2.8) \cdot 10^{-13}$, exploiting 
   this measurement to search for new physics is limited by the knowledge 
   of the electromagnetic fine structure constant, $\alpha$. Inserting independent 
   measurements of $\alpha$ from atom recoil 
   analyses~\cite{Clade:2006zz,Cadoret:2008st,PhysRevA.74.052109,Gerginov:2006zz}, 
   effectively reduces the above accuracy by a factor of 20.
} 
while the tau $g-2$ has not yet been measured to significant precision. 
For recent and very thorough muon $g-2$ reviews, see 
Refs.~\cite{Miller:2007kk,Jegerlehner:2009ry}.

\subsection{Experimental result}

The E821 experiment at Brookhaven National Lab (BNL), USA, studied the precession of 
$\mu^{+}$ and $\mu^{-}$ in a uniform external magnetic field perpendicular to the 
muon spin and orbit plane as they circulated in a cyclotron. At a {\em magic} muon 
momentum of $3.09\:\gev$ (or magic $\gamma=29.3$), 
and negligible $\mu$ EDM, the observed difference 
between spin precession frequency and cyclotron frequency, $\omega_a$ is independent 
of the focusing electric field and is given by $\omega_a=\amu B e/m_\mu c$. 
Following a doubly blinded analysis strategy, E821 found the charge averaged
result~\cite{Bennett:2002jb,Bennett:2004pv,Bennett:2006fi}
\beq
\label{eq:amu_exp_num}
    a^{\rm exp}_{\mu}=(11\,659\,208.9 \pm 5.4 \pm 3.3) \cdot 10^{-10}\,,
\eeq
where the first error is statistical and the second systematic. This result represents 
about a factor of 14 improvement over the classic CERN experiments of the 
1970's~\cite{Bailey:1978mn}.

\subsection{Standard Model prediction}

\begin{figure}[t]
\begin{center}
\includegraphics[width=1\columnwidth]{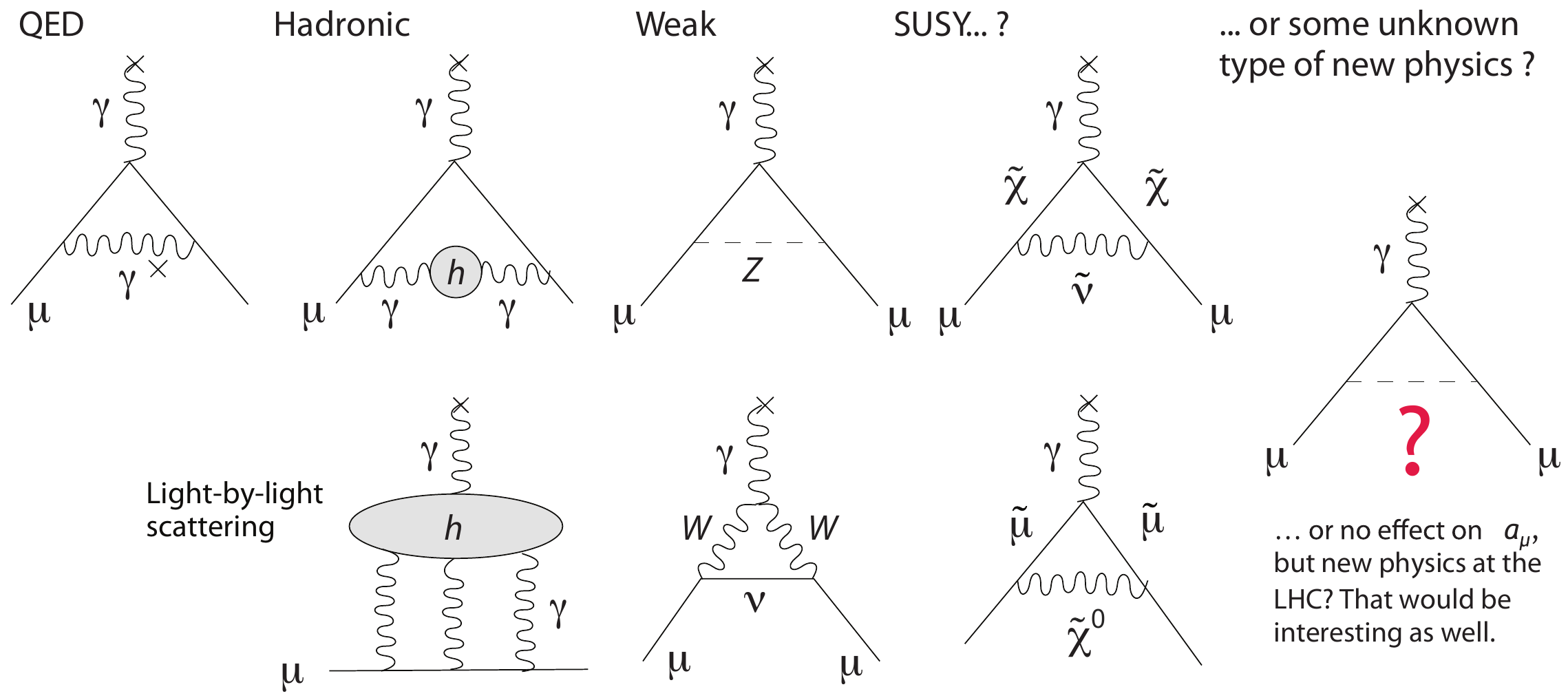}
\end{center}
\vspace{-0.2cm}
\caption{\label{fig:g2-feynman}
        Feynman graphs contributing to $a^{\rm SM}_\mu$ and possible new 
        physics processes.}
\end{figure}
The SM prediction for $a^{\rm SM}_{\mu}$ is conveniently divided into three parts 
(see Fig.~\ref{fig:g2-feynman} for representative Feynman diagrams)
\beq
\label{eq:amusm}
    a^{\rm SM}_{\mu}=a^{\rm QED}_{\mu}+ a^{\rm EW}_{\mu} + \amuhad~.
\eeq
The QED part includes all photonic and leptonic $(e, \mu,\tau)$ loops
starting with the classic $\alpha/2\pi$ Schwinger contribution~\cite{Schwinger:1948iu}. 
It has been computed through 4 loops and estimated at the 5-loop
level~\cite{Kinoshita:2005zr,Aoyama:2007dv,Kinoshita:2004wi,Kinoshita:2005sm,Kataev:2006gx,Kataev:2006yh,Passera:2004bj}.
Employing $\alpha^{-1} = 137.035999084 \pm 51$, 
determined~\cite{Kinoshita:2005zr,Aoyama:2007dv,Kinoshita:2004wi,Kinoshita:2005sm,Kataev:2006gx,Kataev:2006yh,Passera:2004bj,Gabrielse:2006gg,Hanneke:2008tm}
from the electron $a_{e}$ measurement, leads~to
$a^{\rm QED}_{\mu} = (116\,584\,718.09 \pm 0.15) \cdot 10^{-11}$,
where the error results from uncertainties in the coefficients of the 
perturbative series and in $\alpha$. 
Loop contributions involving heavy $W^{\pm},Z$ or Higgs particles are
collectively labelled as $a^{\rm EW}_{\mu}$. They are suppressed by at least 
a factor of $(\alpha/\pi)\cdot(m_{\mu}/m_{W})^2\simeq4\cdot 10^{-9}$.
At 1-loop order one finds 
$a^{\rm EW}_{\mu}[\hbox{1-loop}] = 194.8 \cdot 10^{-11}$~\cite{Jackiw:1972jz}.
Two-loop corrections are found to be relatively large and negative 
$a^{\rm EW}_{\mu}[\hbox{2-loop}] = (-40.7 \pm 1.0 \pm 1.8) \cdot 10^{-11}$~\cite{Czarnecki:2002nt,Heinemeyer:2004yq,Gribouk:2005ee,Czarnecki:1995sz,Czarnecki:1995wq,Peris:1995bb,Kukhto:1992qv},
where the errors stem from quark triangle loops and the assumed Higgs
mass range between 100 and $500\:\gev$. The 3-loop leading logarithms are 
negligible~\cite{Czarnecki:2002nt,Degrassi:1998es}, ${\cal O}(10^{-12})$,
implying for the total weak contribution
$a^{\rm EW}_{\mu} = (154 \pm 1 \pm 2) \cdot 10^{-11}$.

\subsection{Hadronic contribution}

Hadronic  loop contributions to $a^{\rm SM}_{\mu}$
give rise to its main theoretical uncertainties. At present, those
effects are not calculable from first principles, but such an approach,
at least partially, may become possible as lattice QCD matures 
(\cf~\cite{Feng:2011zk} for a lattice calculation using two-flavour QCD 
which results, however, in an underestimate of $a^{{\rm SM},n_{\!f}=2}_{\mu}$).
Instead, one currently relies on a dispersion relation approach to evaluate 
the lowest-order hadronic vacuum polarisation contribution, $\amuhadLO$, 
from corresponding cross section measurements and 
predictions~\cite{Bouchiat:1957zz,Gourdin:1969dm}
\beq
\label{eq:amu_had_lo}
 \amuhadLO =
           \frac{1}{3}\Biggl(\frac{\alpha}{\pi}\Biggr)^{\!2}\!
           \int\limits_{m_\pi^2}^\infty\!\!ds\,\frac{K(s)}{s}R^{(0)}(s)\,,
\eeq
where $R^{(0)}(s)$ denotes the ratio of the bare cross section for $e^+e^-$ 
annihilation into hadrons to the pointlike muon-pair cross section at 
centre-of-mass energy $\sqrt{s}$. The function $K(s)\sim1/s$ gives a 
strong weight to the low-energy part of the integral~\cite{Brodsky:1967sr}, 
such that $\amuhadLO$ is dominated by the $\rho^0(770)\to\pi^+\pi^-$ resonance.

There has been a huge, twenty-years long effort by experimentalists and theorists 
to reduce the error on $\amuhadLO$. It featured improved \ee cross section 
data from the Novosibirsk, Russia accelerator facilities, more use of 
perturbative QCD to replace inaccurate data points, the development of the 
{\em radiative return} technique which allows to exploit cross-section data from 
the high-luminosity $\Phi$ and $B$ factories, and the use of precise $\tau$ 
hadronic spectral functions by applying isospin symmetry. The computation of 
the integral~(\ref{eq:amu_had_lo}) proceeds in different steps. Exclusive
cross-section measurements are summed up to obtain $R^{(0)}(s)$ for $\sqrt{s}\le1.8\:\gev$.
Unmeasured modes are estimated from measured ones using isospin relations. 
Perturbative QCD can be used to predict $R^{(0)}(s)$ away from the quark thresholds in the 
quark-antiquark continuum. Comparisons of the QCD predictions with inclusive data, 
where available, exhibit good agreement with theory. Helpful here is the so-called
global quark-hadron duality approximation, which uses the fact that essentially
perturbative QCD can be used to predict integrals over hadronic spectral 
functions if a large enough spectrum is integrated over and if the endpoint 
of the integral is not too far from the continuum regime. The same concept 
is used to extract \as from $\tau$ decays. Experimental data are used for the 
evaluation of~(\ref{eq:amu_had_lo}) in the charm anti-charm resonance region 
beyond the opening of the $D\Dbar$ threshold. 
\begin{figure}[t]
\begin{center}
\includegraphics[width=0.75\columnwidth]{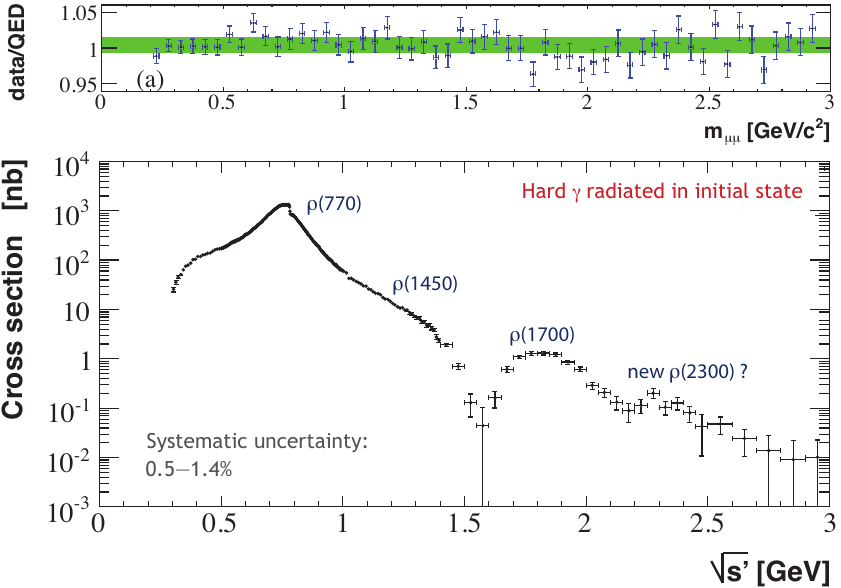}
\end{center}
\vspace{-0.2cm}
\caption{\label{fig:g2-rhobabar2}
        Top panel: ratio of measured cross section for $\ee\to\mu^+\mu^-(\gamma)$
        to the NLO QED prediction. The band is the result of a fit. It includes
        statistical and systematic errors.
        Bottom panel: measured cross section for $\ee\to\pi^+\pi^-(\gamma)$ 
        versus the effective centre-of-mass energy. The data have been unfolded 
        from detector effects. The various known $\rho$ resonances as well as the 
        $\rho$--$\omega$ mixing edge are clearly visible. Figure taken 
        from~\cite{:2009fg} (modified).}
\end{figure}

A significant improvement in the knowledge of the low-energy \ee cross sections
was obtained by realising~\cite{Arbuzov:1998te,Binner:1999bt} that the high
luminosity available at the modern $\Phi$ and $B$ factories made it possible to 
exploit events with hard initial state photon radiation to measure the cross section
of the process $\ee\to {\rm hadrons}(s^\prime)$ at any energy $\sqrt{s^\prime}$ below
$\sqrt{s}$, the centre-of-mass energy of the experiment. It required the measurement 
of the corresponding radiative process, where the photon is emitted by the colliding
electrons so that $s^\prime=s(1-2E_\gamma^\star/\sqrt{s})$, where $E_\gamma^\star$ is 
the centre-of-mass energy of the radiated photon. Figure~\ref{fig:g2-rhobabar2} 
(bottom) shows the $\ee\to\pi^+\pi^-(\gamma)$ cross section measured by BABAR using 
this technique~\cite{:2009fg}. The measurement is systematics dominated, with errors 
ranging between $0.5$ and $1.4\%$. The top panel shows the ratio of the 
$\ee\to\mu^+\mu^-(\gamma)$ cross section between data and the prediction from 
QED. Good agreement is observed, which demonstrates that all relevant efficencies 
and acceptance requirements (trigger, photon, tracking, particle identification, 
kinematic fit, acceptance) are understood. The comparison of different cross section 
data~\cite{Davier:2010nc} revealed a discrepancy between the most precise 
measurements by BABAR and KLOE~\cite{Ambrosino:2010bv,:2008en}. In summer 2011,
KLOE presented a new preliminary measurement~\cite{kloeeps}  of the two-pion cross 
section employing, as was done by BABAR, the $\pi\pi(\gamma) / \mu\mu(\gamma)$
ratio for which radiation, luminosity and vacuum polarisation corrections, as 
well as several acceptance-related systematic uncertainties cancel. The ratio 
crucially relies on a well understood $\pi/\mu$ separation in the 
detector. The analysis used $239\:{\rm pb}^{-1}$ of data, giving $3.4$ ($0.9$)
million $\pi\pi(\gamma)$ ($\mu\mu(\gamma)$) events, and achieved an overall 
systematic precision of $1\%$. Good agreement with the previous KLOE results
is found so that the observed discrepancy with BABAR is corroborated.
\begin{figure}[t]
\begin{center}
\includegraphics[width=0.48\columnwidth]{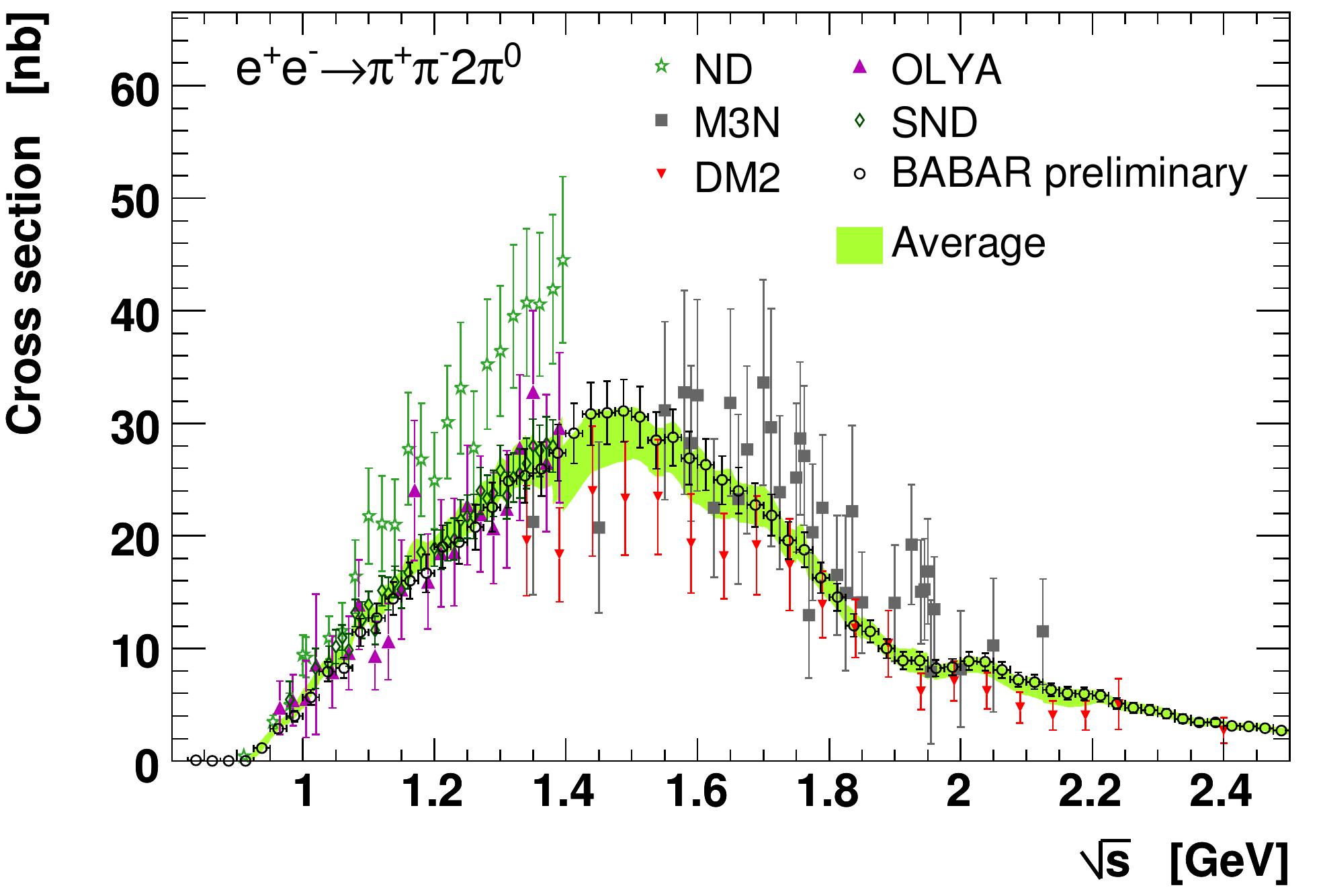}
\includegraphics[width=0.48\columnwidth]{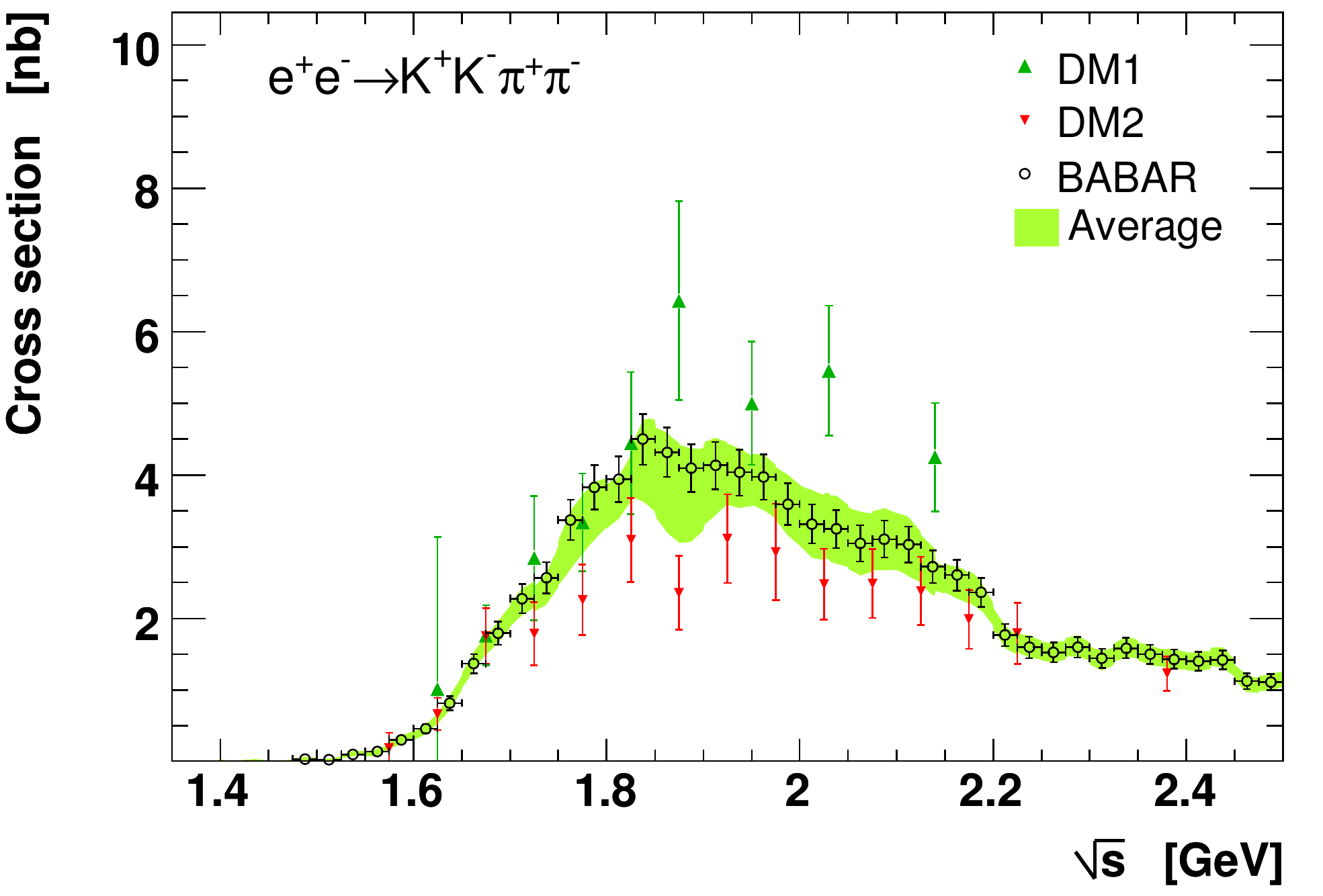}
\end{center}
\vspace{-0.2cm}
\caption{\label{fig:g2-babar}
       Cross section data for the \ee final states $\pi^+\pi^-2\pi^0$ (left) and 
       $\Kp\Km\pp$ (right). The BABAR data~\cite{babar-2pi2pi0,Aubert:2005eg}
       improve the precision and resolve inconsistencies among earlier data sets. 
       Figure taken from~\cite{Davier:2010nc}.}
\end{figure}

Precise BABAR data available for several higher multiplicity modes with and 
without kaons help to discriminate between older, less precise and sometimes 
contradicting measurements. Figure~\ref{fig:g2-babar} shows as an example the 
cross section measurements and their averages for the channels $\ee\to\pi^+\pi^-2\pi^0$
and $\ee\to\Kp\Km\pp$. In several occurrences the older measurements 
overestimate the cross sections in comparison with BABAR, which contributes to 
the reduction in the newer evaluations of the hadronic loop effects. The BABAR
data also greatly help to improve the estimates of unmeasured modes via isospin 
constraints from measured ones~\cite{Davier:2010nc}.

Summing the contributions from all the available $\sigma(e^{+}e^{-} \to{\rm hadrons})$ 
modes and those evaluated with perturbative QCD, the integral~(\ref{eq:amu_had_lo}) 
gives~\cite{Davier:2010nc}
\beq
\label{eq:amu_had_lo_num_ee}
   \amuhadLO = (6\,923 \pm 42 \pm 3) \cdot10^{-11}\,,
\eeq
where the first error is experimental (dominated by systematic uncertainties), 
and the second due to perturbative QCD. New multi-hadron data from the BABAR 
experiment have increased the constraints on unmeasured exclusive final states 
and led to a small reduction in the hadronic contribution compared to previous
evaluations.

Alternatively, one can use precise vector spectral functions from
$\tau \to \nu_{\tau}+{\rm hadrons}$ decays~\cite{Alemany:1997tn} that are
related to isovector $e^{+}e^{-} \to{\rm hadrons}$ cross sections
by isospin symmetry. The $\tau$ data are subject to very different 
systematic uncertainties compared to the \ee data, thus providing a 
valuable cross check. Replacing $e^{+}e^{-}$ data in the two-pion 
and four-pion channels by the corresponding isospin-transformed $\tau$ data, 
and applying isospin-violating corrections (QED radiative corrections, 
charged versus neutral meson mass splitting and electromagnetic
decays, giving $\Delta\amuhadLO=(-3.2 \pm 0.4)\%$) one finds~\cite{Davier:2010nc}
\beq
\label{eq:amu_had_lo_num_tau}
 \amuhadLO = (7\,015 \pm 42 \pm 19 \pm 3) \cdot 10^{-11} \ (\tau)\,,
\eeq
where the first error is experimental, the second estimates the uncertainty in
the isospin-breaking corrections applied to the $\tau$ data, and the third error
is due to perturbative QCD.
The current discrepancy between the $e^{+}e^{-}$ and $\tau$-based determinations
of $\amuhadLO$ has been reduced to $1.8\sigma$ with respect to earlier 
evaluations. New $e^+e^-$ and $\tau$ data from the {\em B}-factory experiments 
BABAR and Belle have increased the experimental information. Reevaluated 
isospin-breaking corrections have also contributed to this 
improvement~\cite{Davier:2009ag}. BABAR reported good agreement with 
the $\tau$ data in the most important two-pion channel~\cite{:2009fg}.
The remaining discrepancy with the older $e^{+}e^{-}$ and $\tau$ datasets
may be indicative of problems with one or both data sets. It may also 
suggest the need for additional isospin-violating corrections to the $\tau$ 
data.

Higher order, ${\cal O}(\alpha^{3})$, hadronic contributions are obtained from 
dispersion relations using the same 
$e^{+}e^{-} \to{\rm hadrons}$ data~\cite{Alemany:1997tn,Hagiwara:2011af,Krause:1996rf}, 
giving $\amuhadHODisp=(-98.4\pm0.6)\cdot10^{-11}$, along with model-dependent estimates
of the hadronic light-by-light scattering contribution~\cite{lbls}, $\amuhadHOLBLS$, motivated by
large-$N_C$ QCD~\cite{Bijnens:2007pz,Prades:2009tw,Melnikov:2003xd,deRafael:1993za}.
Following~\cite{Prades:2009tw}, one finds for the sum of the two terms
$\amuhadHO = (7\pm26) \cdot 10^{-11}$, where the error is dominated by the 
light-by-light contribution.

\begin{figure}[t]
\begin{center}
\includegraphics[width=0.6\columnwidth]{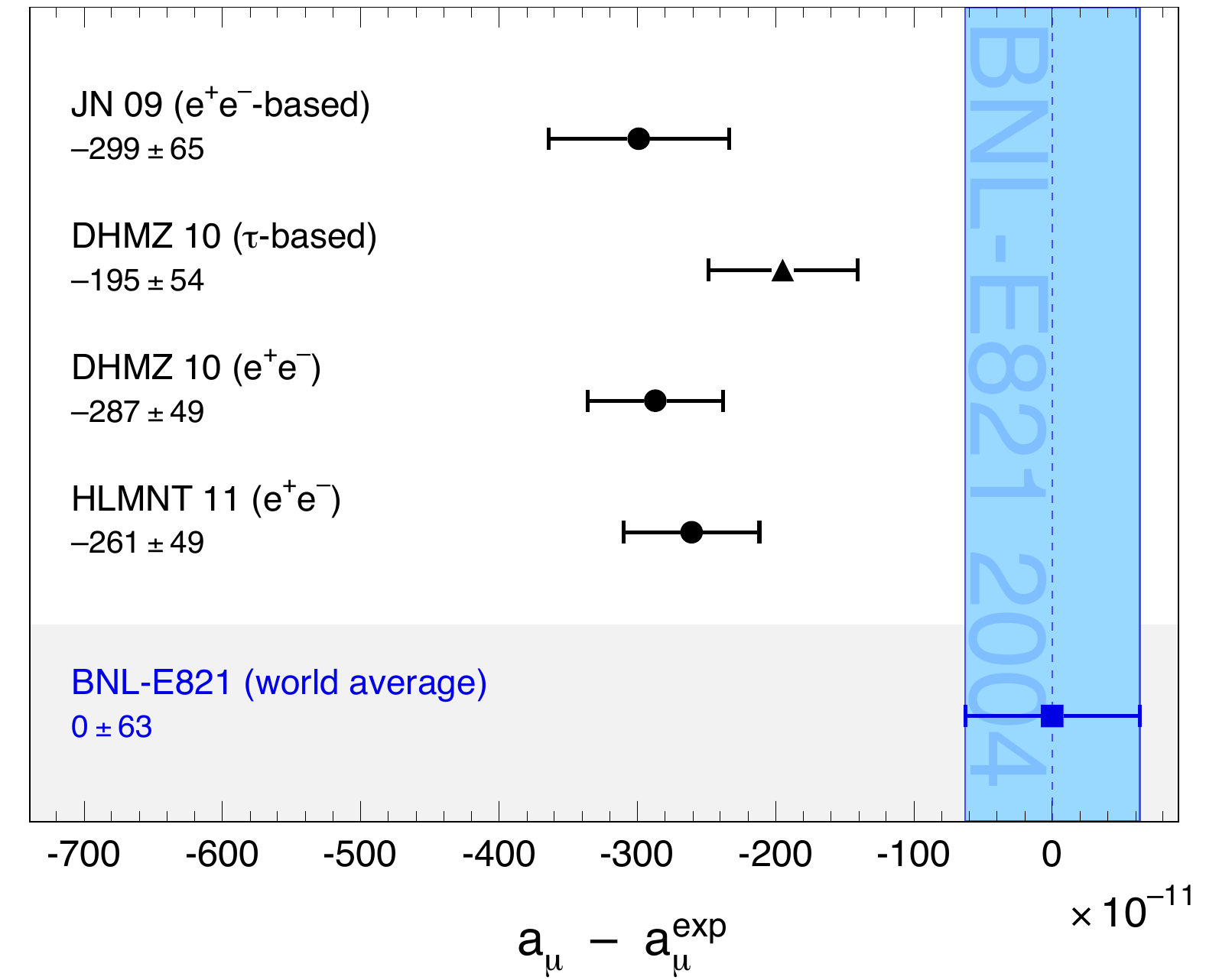}
\end{center}
\caption{\label{fig:amures}
        Compilation of recently published results for $a_\mu$
        (in units of $10^{-11}$),
        subtracted by the central value of the experimental
        average~(\ref{eq:amu_exp_num}).
        The shaded band indicates the experimental error.
        The SM predictions are taken from: 
        JN~\cite{Jegerlehner:2009ry}, 
        DHMZ~\cite{Davier:2010nc}, and
        HMNT~\cite{Hagiwara:2011af}.
        Note that the quoted errors do not
        include the uncertainty on the subtracted experimental
        value. Figure from~\cite{hoeckermarciano}.}
\end{figure}

Adding all SM contributions gives the \ee data based SM prediction 
\beq
   a^{\rm SM}_{\mu} = (116\,591\,802 \pm 2 \pm 42 \pm 26) \cdot 10^{-11}\,,
\eeq
where the errors are due to the electroweak, lowest-order hadronic, 
and higher-order hadronic contributions, respectively.
The difference between experiment and theory,
$\Delta a_{\mu} = a^{\rm exp}_{\mu} - a^{\rm SM}_{\mu}=(287\pm 63 \pm 49)\cdot 10^{-11}$
(with all errors combined in quadrature), represents an interesting
but not yet conclusive discrepancy of approximately $3.6\sigma$. All the recent 
estimates for the hadronic contribution compiled in Fig.~\ref{fig:amures} 
exhibit similar discrepancies. Switching to $\tau$ data reduces the 
discrepancy to $2.4\sigma$, assuming the isospin-violating corrections
are under control within the estimated uncertainties.

An alternate interpretation is that $\Delta a_{\mu}$ may be a new physics
signal with supersymmetric particle loops. Such a scenario is quite natural, 
since generically, supersymmetric models predict~\cite{Czarnecki:2001pv} an 
additional contribution to $a_\mu^{\rm SM}$ of order
\beq
 a_{\mu}^{\rm SUSY} \simeq \pm\, 130 \cdot 10^{-11}\cdot
        \left(\frac{100\;{\rm GeV}}{m_{\rm SUSY}}\right)^{\!\!2}
        \tan\!\beta\,,
\eeq
where $m_{\rm SUSY}$ is a representative supersymmetric mass scale,
and $\tan\!\beta \simeq 3$--$40$ is a potential enhancement factor.
Supersymmetric particles in the mass range $100$--$500\;{\rm GeV}$
could be the source of the deviation $\Delta a_{\mu}$. If so, those
particles could be directly observed at the LHC. The experiments ATLAS 
and CMS at the LHC are searching for supersymmetry in a large variety of 
signatures already pushing the limits for squarks and gluinos beyond the \tev 
scale for the most sensitive searches in events with jets and missing transverse
energy~\cite{Aad:2011ib,Chatrchyan:2011zy}. These limits, however, do not 
directly affect the loops contributing to $a_{\mu}^{\rm SUSY}$, which are dominated 
by virtual gaugino and slepton exchange (\cf Fig.~\ref{fig:g2-feynman}), or only in 
very constrained models~\cite{Buchmueller:2011ki}. 

New physics effects~\cite{Czarnecki:2001pv} other than supersymmetry could 
also explain a non-vanishing $\Delta a_{\mu}$. A recent popular scenario 
involves the ``dark photon'', a relatively light hypothetical vector 
boson from the dark matter sector that couples to our world of particle 
physics through mixing with the ordinary photon~\cite{Pospelov:2008zw,TuckerSmith:2010ra}.
As a result, it couples to ordinary charged particles with strength 
$\varepsilon\cdot e$ and gives rise to an additional muon anomalous magnetic 
moment contribution
$a_\mu^{\rm dark\;photon} = (\alpha/2\pi)\cdot\varepsilon^2 F(m_V/m_\mu)$,
where $F(x)$ is a monotonously falling function with $F(0.1)=0.8$, $F(1)=0.2$.
For values of
$\varepsilon\sim1$--$2\cdot10^{-3}$ and $m_V\sim10$--100$\,$MeV, the dark photon, 
which was originally motivated by cosmology, can provide a viable solution 
to the muon $g-2$ discrepancy. Searches for the dark photon in that mass 
range are currently underway at Jefferson Lab, USA, and MAMI in Mainz, Germany.

Despite the significant experimental and theoretical progress reported here,
the situation of the muon $g-2$ stays inconclusive. Proposals for new experiments 
based on different experimental approaches exist at FNAL~\cite{fnal_g-2} and 
J-PARC~\cite{jparc_g-2}. The FNAL proposal continues the established 
magic-$\gamma$ technique, but with higher proton rate than at BNL and less protons 
per bunch, a $900\:$m pion decay line (BNL: $80\:$m) leading to a smaller pion flash 
at the muon ring injection, zero-degree muons giving a $5$--$10$ times larger 
muon yield per proton, and $5$--$10$ times as many muons stored per hour than
at the BNL experiment. In addition, improved detectors against signal pileup, 
new electronics, and a better shimming to reduce $B$-field variations should 
lead to a $2.5$ ($3$) times smaller systematic error on the $B$-field ($\omega_a$).
The experiment targets a $20$ times better statistical error, 
and a final systematic error of $1.6\cdot10^{-10}$. The J-PARC proposal anticipates 
similar precision without a focusing electric field and hence no need to choose
a particular magic $\gamma$. Ultra-slow muons ($p_\mu\sim2.3\:$keV), generated 
from laser-ionised muonium atoms, are used which have a transverse momentum 
dispersion that is significantly smaller than the longitudinal momentum, allowing 
the beam to circulate in the storage ring without focusing field. 

\section{Conclusions}

All of us are passionately following the LHC experiments in their direct 
searches for electroweak symmetry breaking remnants and new physics at the 
TeV-scale energy frontier. Probing new physics orders of magnitude beyond 
that scale, and helping to decipher possible TeV-scale new physics discovered
at the LHC requires to work hard on the intensity and precision frontiers.

Charged leptons offer an important spectrum of possibilities. Charged-lepton flavour
violation and electric dipole moment measurements have Standard Model free signals,
and current experiments and mature proposals promise orders of magnitude sensitivity 
improvement over the state of the art. The muon anomalous magnetic moment may already 
witness a deviation from the Standard Model. New physics models often strongly 
correlate all these sectors of paricle physics.

\vspace{0.5cm}
{\small 
\noindent
{\bf Acknowledgements} --- I am grateful to Alessandro Baldini, Swagato Banerjee,
Robert Bernstein, Michel Davier, Tim Gershon, Herve Hiu Fai
Choi, Yoshitaka Kuno, Alberto Lusiani, Bogdan Malaescu, James
Miller, Toshinori Mori, Steven Robertson, Karim Trabelsi, Graziano
Venanzoni and Zhiqing Zhang for their help with the preparation
of the talk. It is a pleasure to thank the organisers
and the international advisory board of LP11 for giving me the opportunity 
to talk at this important conference and to visit the extraordinarily 
interesting and dynamic city of Mumbai.
}

\bibliographystyle{pramana}
\bibliography{lp11_bibliography}{}

\begin{thebibliography}{100}
\providecommand{\url}[1]{{\tt #1}}
\providecommand{\urlprefix}{URL }
\providecommand{\bibinfo}[2]{#2}
\providecommand{\eprint}[2][]{\url{#2}}

\bibitem{Glashow:1970gm}
\bibinfo{author}{S.L. Glashow}, \bibinfo{author}{J.~Iliopoulos} and
  \bibinfo{author}{L.~Maiani}, {\em \bibinfo{journal}{Phys.Rev.}\/} {\bf
  \bibinfo{volume}{D2}}, \bibinfo{pages}{1285} (\bibinfo{year}{1970}).

\bibitem{LHCb:2011ac}
\bibinfo{author}{LHCb Collaboration}  (\bibinfo{year}{2011}).
  \eprint{1112.1600}.

\bibitem{PhysRevLett.107.191802}
\bibinfo{author}{CMS Collaboration}, {\em \bibinfo{journal}{Phys. Rev.
  Lett.}\/} {\bf \bibinfo{volume}{107}}, \bibinfo{pages}{191802}
  (\bibinfo{year}{2011}).

\bibitem{na62}
 \bibinfo{note}{NA62 Collaboration, 2011 Status Report to the CERN SPSC},
  \urlprefix\url{http://na62.web.cern.ch/na62/Documents/SPSC-SR-083.pdf}.

\bibitem{Marciano:2008zz}
\bibinfo{author}{W.J. Marciano}, \bibinfo{author}{T.~Mori} and
  \bibinfo{author}{J.M. Roney}, {\em
  \bibinfo{journal}{Ann.Rev.Nucl.Part.Sci.}\/} {\bf \bibinfo{volume}{58}},
  \bibinfo{pages}{315} (\bibinfo{year}{2008}).

\bibitem{Asner:2010qj}
\bibinfo{author}{Heavy Flavor~Averaging Group}  (\bibinfo{year}{2010}).
  \eprint{1010.1589},
  \urlprefix\url{http://www.slac.stanford.edu/xorg/hfag/tau/index.html}.

\bibitem{PhysRevLett.9.36}
\bibinfo{author}{G.~Danby}, \bibinfo{author}{J-M. Gaillard},
  \bibinfo{author}{K.~Goulianos}, \bibinfo{author}{L.M. Lederman},
  \bibinfo{author}{N.~Mistry}, \bibinfo{author}{M.~Schwartz} and
  \bibinfo{author}{J.~Steinberger}, {\em \bibinfo{journal}{Phys. Rev. Lett.}\/}
  {\bf \bibinfo{volume}{9}}, \bibinfo{pages}{36} (\bibinfo{year}{1962}).

\bibitem{Feinberg:1958zzb}
\bibinfo{author}{G.~Feinberg}, {\em \bibinfo{journal}{Phys.Rev.}\/} {\bf
  \bibinfo{volume}{110}}, \bibinfo{pages}{1482} (\bibinfo{year}{1958}).

\bibitem{Ahmed:2001eh}
\bibinfo{author}{M.~Ahmed} {\em et~al.\/} (\bibinfo{collaboration}{MEGA
  Collaboration}), {\em \bibinfo{journal}{Phys.Rev.}\/} {\bf
  \bibinfo{volume}{D65}}, \bibinfo{pages}{112002} (\bibinfo{year}{2002}).
  \eprint{hep-ex/0111030}.

\bibitem{Kuno:1999jp}
\bibinfo{author}{Y.~Kuno} and \bibinfo{author}{Y.~Okada}, {\em
  \bibinfo{journal}{Rev.Mod.Phys.}\/} {\bf \bibinfo{volume}{73}},
  \bibinfo{pages}{151} (\bibinfo{year}{2001}). \eprint{hep-ph/9909265}.

\bibitem{Raidal:2008jk}
\bibinfo{author}{M.~Raidal} {\em et~al.\/}, {\em
  \bibinfo{journal}{120Eur.Phys.J.}\/} {\bf \bibinfo{volume}{C57}},
  \bibinfo{pages}{13} (\bibinfo{year}{2008}). \eprint{0801.1826}.

\bibitem{Barbier:2004ez}
\bibinfo{author}{R.~Barbier}, \bibinfo{author}{C.~Berat},
  \bibinfo{author}{M.~Besancon}, \bibinfo{author}{M.~Chemtob},
  \bibinfo{author}{A.~Deandrea} {\em et~al.\/}, {\em
  \bibinfo{journal}{Phys.Rept.}\/} {\bf \bibinfo{volume}{420}},
  \bibinfo{pages}{1} (\bibinfo{year}{2005}). \eprint{hep-ph/0406039}.

\bibitem{meg1}
 \bibinfo{note}{A. Baldini, T. Mori et al., {\em The MEG experiment: search for
  the $\mu\to e\gamma$ decay at PSI}}, \urlprefix\url{http://meg.psi.ch/docs}.

\bibitem{Adam:2009ci}
\bibinfo{author}{J.~Adam} {\em et~al.\/} (\bibinfo{collaboration}{MEG
  collaboration}), {\em \bibinfo{journal}{Nucl.Phys.}\/} {\bf
  \bibinfo{volume}{B834}}, \bibinfo{pages}{1} (\bibinfo{year}{2010}).
  \eprint{0908.2594}.

\bibitem{Sawada:2010zz}
\bibinfo{author}{R.~Sawada} (\bibinfo{collaboration}{MEG Collaboration}), {\em
  \bibinfo{journal}{PoS}\/} {\bf \bibinfo{volume}{ICHEP2010}},
  \bibinfo{pages}{263} (\bibinfo{year}{2010}).

\bibitem{Adam:2011ch}
\bibinfo{author}{MEG collaboration}, {\em \bibinfo{journal}{Phys.Rev.Lett.}\/}
  {\bf \bibinfo{volume}{107}}, \bibinfo{pages}{171801} (\bibinfo{year}{2011}).
  \eprint{1107.5547}.

\bibitem{Calibbi:2006nq}
\bibinfo{author}{L.~Calibbi}, \bibinfo{author}{A.~Faccia},
  \bibinfo{author}{A.~Masiero} and \bibinfo{author}{S.K. Vempati}, {\em
  \bibinfo{journal}{Phys.Rev.}\/} {\bf \bibinfo{volume}{D74}},
  \bibinfo{pages}{116002} (\bibinfo{year}{2006}). \eprint{hep-ph/0605139}.

\bibitem{Masiero:2002jn}
\bibinfo{author}{A.~Masiero}, \bibinfo{author}{S.~K. Vempati} and
  \bibinfo{author}{O.~Vives}, {\em \bibinfo{journal}{Nucl.Phys.}\/} {\bf
  \bibinfo{volume}{B649}}, \bibinfo{pages}{189} (\bibinfo{year}{2003}).
  \eprint{hep-ph/0209303}.

\bibitem{Cirigliano:2009bz}
\bibinfo{author}{V.~Cirigliano}, \bibinfo{author}{R.~Kitano},
  \bibinfo{author}{Y.~Okada} and \bibinfo{author}{P.~Tuzon}, {\em
  \bibinfo{journal}{Phys.Rev.}\/} {\bf \bibinfo{volume}{D80}},
  \bibinfo{pages}{013002} (\bibinfo{year}{2009}). \eprint{0904.0957}.

\bibitem{Kitano:2003wn}
\bibinfo{author}{R.~Kitano}, \bibinfo{author}{M.~Koike},
  \bibinfo{author}{S.~Komine} and \bibinfo{author}{Y.~Okada}, {\em
  \bibinfo{journal}{Phys.Lett.}\/} {\bf \bibinfo{volume}{B575}},
  \bibinfo{pages}{300} (\bibinfo{year}{2003}). \eprint{hep-ph/0308021}.

\bibitem{Czarnecki:1998iz}
\bibinfo{author}{A.~Czarnecki}, \bibinfo{author}{W.J. Marciano} and
  \bibinfo{author}{K.~Melnikov}, {\em \bibinfo{journal}{AIP Conf.Proc.}\/} {\bf
  \bibinfo{volume}{435}}, \bibinfo{pages}{409} (\bibinfo{year}{1998}).
  \eprint{hep-ph/9801218}.

\bibitem{Kitano:2002mt}
\bibinfo{author}{R.~Kitano}, \bibinfo{author}{M.~Koike} and
  \bibinfo{author}{Y.~Okada}, {\em \bibinfo{journal}{Phys.Rev.}\/} {\bf
  \bibinfo{volume}{D66}}, \bibinfo{pages}{096002} (\bibinfo{year}{2002}).
  \eprint{hep-ph/0203110}.

\bibitem{Bertl:2006up}
\bibinfo{author}{W.H. Bertl} {\em et~al.\/} (\bibinfo{collaboration}{SINDRUM II
  Collaboration}), {\em \bibinfo{journal}{Eur.Phys.J.}\/} {\bf
  \bibinfo{volume}{C47}}, \bibinfo{pages}{337} (\bibinfo{year}{2006}).

\bibitem{mu2e}
 \bibinfo{note}{See, \eg, R. Bernstein, Pittsburgh Seminar, Feb 2011},
  \urlprefix\url{http://mu2e.fnal.gov}.

\bibitem{comet}
 \bibinfo{note}{COMET Collaboration, KEK Report 2009-10 (TDR)}.

\bibitem{projectx}
 \bibinfo{note}{Project-X, FNAL}, \urlprefix\url{http://projectx.fnal.gov/}.

\bibitem{prism}
 \bibinfo{note}{J. Pasternak {\em et al.}, Proc. Int. Particle Accel. Conf.
  IPAC10, Kyoto, Japan 2010; A. Sato {\em et al.}, Proc. EPAC2006, p2508,
  Edinburgh 2006}.

\bibitem{Gonderinger:2010yn}
\bibinfo{author}{Matthew Gonderinger} and \bibinfo{author}{Michael~J.
  Ramsey-Musolf}, {\em \bibinfo{journal}{JHEP}\/} {\bf \bibinfo{volume}{1011}},
  \bibinfo{pages}{045} (\bibinfo{year}{2010}). \eprint{1006.5063}.

\bibitem{:2009tk}
\bibinfo{author}{BABAR Collaboration}, {\em
  \bibinfo{journal}{Phys.Rev.Lett.}\/} {\bf \bibinfo{volume}{104}},
  \bibinfo{pages}{021802} (\bibinfo{year}{2010}). \eprint{0908.2381}.

\bibitem{Miyazaki:2011xe}
\bibinfo{author}{Y.~Miyazaki} {\em et~al.\/}, {\em \bibinfo{journal}{Phys.
  Lett.}\/} {\bf \bibinfo{volume}{B699}}, \bibinfo{pages}{251}
  (\bibinfo{year}{2011}). \eprint{1101.0755}.

\bibitem{Bona:2007qt}
\bibinfo{author}{SuperB Collaboration}  (\bibinfo{year}{2007}).
  \eprint{0709.0451}.

\bibitem{Abe:2010sj}
\bibinfo{author}{Belle~II Collaboration}  (\bibinfo{year}{2010}).
  \bibinfo{note}{Long author list - awaiting processing}, \eprint{1011.0352}.

\bibitem{1742-6596-171-1-012079}
\bibinfo{author}{K~Hayasaka}, {\em \bibinfo{journal}{Journal of Physics:
  Conference Series}\/} {\bf \bibinfo{volume}{171}}, \bibinfo{number}{1},
  \bibinfo{pages}{012079} (\bibinfo{year}{2009}).

\bibitem{Abe:2011sj}
\bibinfo{author}{T2K Collaboration}, {\em \bibinfo{journal}{Phys.Rev.Lett.}\/}
  {\bf \bibinfo{volume}{107}}, \bibinfo{pages}{041801} (\bibinfo{year}{2011}).
  \eprint{1106.2822}.

\bibitem{Hisano:2009ae}
\bibinfo{author}{J.~Hisano}, \bibinfo{author}{M.~Nagai},
  \bibinfo{author}{P.~Paradisi} and \bibinfo{author}{Y.~Shimizu}, {\em
  \bibinfo{journal}{JHEP}\/} {\bf \bibinfo{volume}{0912}}, \bibinfo{pages}{030}
  (\bibinfo{year}{2009}). \eprint{0904.2080}.

\bibitem{Lamoreaux:2009zz}
\bibinfo{author}{S.K. Lamoreaux} and \bibinfo{author}{R.~Golub}, {\em
  \bibinfo{journal}{J.Phys.G}\/} {\bf \bibinfo{volume}{G36}},
  \bibinfo{pages}{104002} (\bibinfo{year}{2009}).

\bibitem{Pospelov:2005pr}
\bibinfo{author}{M.~Pospelov} and \bibinfo{author}{A.~Ritz}, {\em
  \bibinfo{journal}{Annals Phys.}\/} {\bf \bibinfo{volume}{318}},
  \bibinfo{pages}{119} (\bibinfo{year}{2005}). \eprint{hep-ph/0504231}.

\bibitem{KirchPanic}
 \bibinfo{note}{K.~Kirch, Talk at PANIC 2011, MIT-Cambridge, USA}.

\bibitem{Zsigmond}
 \bibinfo{note}{G.~Zsigmond, Talk at EPS-HEP 2011, Grenoble, France}.

\bibitem{Baker:2006ts}
\bibinfo{author}{C.A. Baker}, \bibinfo{author}{D.D. Doyle},
  \bibinfo{author}{P.~Geltenbort}, \bibinfo{author}{K.~Green},
  \bibinfo{author}{M.G.D. van~der Grinten} {\em et~al.\/}, {\em
  \bibinfo{journal}{Phys.Rev.Lett.}\/} {\bf \bibinfo{volume}{97}},
  \bibinfo{pages}{131801} (\bibinfo{year}{2006}). \eprint{hep-ex/0602020}.

\bibitem{protonEDM-BNL}
 \bibinfo{note}{Storage Ring EDM Collaboration},
  \urlprefix\url{http://www.bnl.gov/edm/}.

\bibitem{Semertzidis}
 \bibinfo{note}{Y. Semertzidis, Talk at Patras-Axion workshop 2011, Mykonos,
  Greece}.

\bibitem{Regan:2002ta}
\bibinfo{author}{B.C. Regan}, \bibinfo{author}{E.D. Commins},
  \bibinfo{author}{C.J. Schmidt} and \bibinfo{author}{D.~DeMille}, {\em
  \bibinfo{journal}{Phys.Rev.Lett.}\/} {\bf \bibinfo{volume}{88}},
  \bibinfo{pages}{071805} (\bibinfo{year}{2002}).

\bibitem{eedm}
 \bibinfo{note}{Commins, E. D. Electric dipole moments of leptons. In Advances
  in Atomic, Molecular, and Optical Physics Vol. 40, 1–56 (eds Bederson, B.
  and Walther, H.), Academic Press. (1999)}.

\bibitem{Hudson:2011zz}
\bibinfo{author}{J.J. Hudson}, \bibinfo{author}{D.M. Kara},
  \bibinfo{author}{I.J. Smallman}, \bibinfo{author}{B.E. Sauer},
  \bibinfo{author}{M.R. Tarbutt} {\em et~al.\/}, {\em
  \bibinfo{journal}{Nature}\/} {\bf \bibinfo{volume}{473}},
  \bibinfo{pages}{493} (\bibinfo{year}{2011}).

\bibitem{Schiff:1963zz}
\bibinfo{author}{L.I. Schiff}, {\em \bibinfo{journal}{Phys.Rev.}\/} {\bf
  \bibinfo{volume}{132}}, \bibinfo{pages}{2194} (\bibinfo{year}{1963}).

\bibitem{Hudson:2002az}
\bibinfo{author}{J.J Hudson}, \bibinfo{author}{B.E. Sauer},
  \bibinfo{author}{M.R. Tarbutt} and \bibinfo{author}{E.A. Hinds}, {\em
  \bibinfo{journal}{Phys.Rev.Lett.}\/} {\bf \bibinfo{volume}{89}},
  \bibinfo{pages}{023003} (\bibinfo{year}{2002}). \eprint{hep-ex/0202014}.

\bibitem{Schael:2005am}
\bibinfo{author}{ALEPH Collaboration}, {\em \bibinfo{journal}{Phys.Rept.}\/}
  {\bf \bibinfo{volume}{421}}, \bibinfo{pages}{191} (\bibinfo{year}{2005}).
  \eprint{hep-ex/0506072}.

\bibitem{Nakamura:2010zzi}
\bibinfo{author}{K.~Nakamura} {\em et~al.\/} (\bibinfo{collaboration}{Particle
  Data Group}), {\em \bibinfo{journal}{J.Phys.G}\/} {\bf
  \bibinfo{volume}{G37}}, \bibinfo{pages}{075021} (\bibinfo{year}{2010}).

\bibitem{:2005ema}
{\em \bibinfo{journal}{Phys.Rept.}\/} {\bf \bibinfo{volume}{427}},
  \bibinfo{pages}{257} (\bibinfo{year}{2006}). \eprint{hep-ex/0509008}.

\bibitem{Weinberg:1958ut}
\bibinfo{author}{S.~Weinberg}, {\em \bibinfo{journal}{Phys.Rev.}\/} {\bf
  \bibinfo{volume}{112}}, \bibinfo{pages}{1375} (\bibinfo{year}{1958}).

\bibitem{Pich:1987qq}
\bibinfo{author}{A.~Pich}, {\em \bibinfo{journal}{Phys.Lett.}\/} {\bf
  \bibinfo{volume}{B196}}, \bibinfo{pages}{561} (\bibinfo{year}{1987}).

\bibitem{Nussinov:2008gx}
\bibinfo{author}{S.~Nussinov} and \bibinfo{author}{A.~Soffer}, {\em
  \bibinfo{journal}{Phys.Rev.}\/} {\bf \bibinfo{volume}{D78}},
  \bibinfo{pages}{033006} (\bibinfo{year}{2008}). \eprint{0806.3922}.

\bibitem{delAmoSanchez:2010pc}
\bibinfo{author}{BABAR Collaboration}, {\em \bibinfo{journal}{Phys.Rev.}\/}
  {\bf \bibinfo{volume}{D83}}, \bibinfo{pages}{032002} (\bibinfo{year}{2011}).
  \eprint{1011.3917}.

\bibitem{Bigi:2005ts}
\bibinfo{author}{I.I. Bigi} and \bibinfo{author}{A.I. Sanda}, {\em
  \bibinfo{journal}{Phys.Lett.}\/} {\bf \bibinfo{volume}{B625}},
  \bibinfo{pages}{47} (\bibinfo{year}{2005}). \eprint{hep-ph/0506037}.

\bibitem{Grossman:2011zk}
\bibinfo{author}{Y.~Grossman} and \bibinfo{author}{Y~Nir}
  (\bibinfo{year}{2011}). \eprint{1110.3790}.

\bibitem{Kuhn:1996dv}
\bibinfo{author}{J.H. Kuhn} and \bibinfo{author}{E.~Mirkes}, {\em
  \bibinfo{journal}{Phys.Lett.}\/} {\bf \bibinfo{volume}{B398}},
  \bibinfo{pages}{407} (\bibinfo{year}{1997}). \eprint{hep-ph/9609502}.

\bibitem{delAmoSanchez:2011zza}
\bibinfo{author}{BABAR Collaboration}, {\em \bibinfo{journal}{Phys.Rev.}\/}
  {\bf \bibinfo{volume}{D83}}, \bibinfo{pages}{071103} (\bibinfo{year}{2011}).
  \eprint{1011.5477}.

\bibitem{Bischofberger:2011pw}
\bibinfo{author}{Belle Collaboration}, {\em
  \bibinfo{journal}{Phys.Rev.Lett.}\/} {\bf \bibinfo{volume}{107}},
  \bibinfo{pages}{131801} (\bibinfo{year}{2011}). \eprint{1101.0349}.

\bibitem{BABAR:2011aa}
\bibinfo{author}{BABAR Collaboration}  (\bibinfo{year}{2011}).
  \eprint{1109.1527}.

\bibitem{Davier:2005xq}
\bibinfo{author}{M.~Davier}, \bibinfo{author}{A.~Hoecker} and
  \bibinfo{author}{A.~Zhang}, {\em \bibinfo{journal}{Rev.Mod.Phys.}\/} {\bf
  \bibinfo{volume}{78}}, \bibinfo{pages}{1043} (\bibinfo{year}{2006}).
  \eprint{hep-ph/0507078}.

\bibitem{Barate:1997hv}
\bibinfo{author}{ALEPH Collaboration}, {\em \bibinfo{journal}{Z.Phys.}\/} {\bf
  \bibinfo{volume}{C76}}, \bibinfo{pages}{15} (\bibinfo{year}{1997}).

\bibitem{Barate:1998uf}
\bibinfo{author}{ALEPH Collaboration}, {\em \bibinfo{journal}{Eur.Phys.J.}\/}
  {\bf \bibinfo{volume}{C4}}, \bibinfo{pages}{409} (\bibinfo{year}{1998}).

\bibitem{Ackerstaff:1998yj}
\bibinfo{author}{OPAL Collaboration}, {\em \bibinfo{journal}{Eur.Phys.J.}\/}
  {\bf \bibinfo{volume}{C7}}, \bibinfo{pages}{571} (\bibinfo{year}{1999}).
  \eprint{hep-ex/9808019}.

\bibitem{Boito:2011qt}
\bibinfo{author}{Diogo Boito}, \bibinfo{author}{Oscar Cata},
  \bibinfo{author}{Maarten Golterman}, \bibinfo{author}{Matthias Jamin},
  \bibinfo{author}{Kim Maltman} {\em et~al.\/}, {\em
  \bibinfo{journal}{Phys.Rev.}\/} {\bf \bibinfo{volume}{D84}},
  \bibinfo{pages}{113006} (\bibinfo{year}{2011}). \eprint{1110.1127}.

\bibitem{Boito:2011pr}
\bibinfo{author}{D.~Boito}, \bibinfo{author}{O.~Cata},
  \bibinfo{author}{M.~Golterman}, \bibinfo{author}{M.~Jamin},
  \bibinfo{author}{K.~Maltman} {\em et~al.\/}  (\bibinfo{year}{2011}).
  \eprint{1112.4202}.

\bibitem{Baikov:2008jh}
\bibinfo{author}{P.A. Baikov}, \bibinfo{author}{K.G. Chetyrkin} and
  \bibinfo{author}{J.H. Kuhn}, {\em \bibinfo{journal}{Phys.Rev.Lett.}\/} {\bf
  \bibinfo{volume}{101}}, \bibinfo{pages}{012002} (\bibinfo{year}{2008}).
  \eprint{0801.1821}.

\bibitem{Davier:2008sk}
\bibinfo{author}{M.~Davier}, \bibinfo{author}{S.~Descotes-Genon},
  \bibinfo{author}{A.~Hoecker}, \bibinfo{author}{B.~Malaescu} and
  \bibinfo{author}{Z.~Zhang}, {\em \bibinfo{journal}{Eur.Phys.J.}\/} {\bf
  \bibinfo{volume}{C56}}, \bibinfo{pages}{305} (\bibinfo{year}{2008}).
  \eprint{0803.0979}.

\bibitem{Beneke:2008ad}
\bibinfo{author}{M.~Beneke} and \bibinfo{author}{M.~Jamin}, {\em
  \bibinfo{journal}{JHEP}\/} {\bf \bibinfo{volume}{0809}}, \bibinfo{pages}{044}
  (\bibinfo{year}{2008}). \eprint{0806.3156}.

\bibitem{Menke:2009vg}
\bibinfo{author}{A.~Menke}  (\bibinfo{year}{2009}). \eprint{0904.1796}.

\bibitem{Caprini:2011ya}
\bibinfo{author}{I.~Caprini} and \bibinfo{author}{J.~Fischer}, {\em
  \bibinfo{journal}{Phys.Rev.}\/} {\bf \bibinfo{volume}{D84}},
  \bibinfo{pages}{054019} (\bibinfo{year}{2011}). \eprint{1106.5336}.

\bibitem{vanRitbergen:1997va}
\bibinfo{author}{T.~van Ritbergen}, \bibinfo{author}{J.A.M. Vermaseren} and
  \bibinfo{author}{S.A. Larin}, {\em \bibinfo{journal}{Phys.Lett.}\/} {\bf
  \bibinfo{volume}{B400}}, \bibinfo{pages}{379} (\bibinfo{year}{1997}).
  \eprint{hep-ph/9701390}.

\bibitem{Chetyrkin:1997sg}
\bibinfo{author}{K.G. Chetyrkin}, \bibinfo{author}{Bernd~A. Kniehl} and
  \bibinfo{author}{M.~Steinhauser}, {\em \bibinfo{journal}{Phys.Rev.Lett.}\/}
  {\bf \bibinfo{volume}{79}}, \bibinfo{pages}{2184} (\bibinfo{year}{1997}).
  \eprint{hep-ph/9706430}.

\bibitem{Chetyrkin:1997un}
\bibinfo{author}{K.G. Chetyrkin}, \bibinfo{author}{B.A. Kniehl} and
  \bibinfo{author}{M.~Steinhauser}, {\em \bibinfo{journal}{Nucl.Phys.}\/} {\bf
  \bibinfo{volume}{B510}}, \bibinfo{pages}{61} (\bibinfo{year}{1998}).
  \eprint{hep-ph/9708255}.

\bibitem{Rodrigo:1997zd}
\bibinfo{author}{G.~Rodrigo}, \bibinfo{author}{A.~Pich} and
  \bibinfo{author}{A.~Santamaria}, {\em \bibinfo{journal}{Phys.Lett.}\/} {\bf
  \bibinfo{volume}{B424}}, \bibinfo{pages}{367} (\bibinfo{year}{1998}).
  \eprint{hep-ph/9707474}.

\bibitem{Baak:2011ze}
\bibinfo{author}{M.~Baak} {\em et~al.\/}  (\bibinfo{year}{2011}).
  \eprint{1107.0975}.

\bibitem{Bethke:2009jm}
\bibinfo{author}{S.~Bethke}, {\em \bibinfo{journal}{Eur.Phys.J.}\/} {\bf
  \bibinfo{volume}{C64}}, \bibinfo{pages}{689} (\bibinfo{year}{2009}).
  \eprint{0908.1135}.

\bibitem{Gamiz:2004ar}
\bibinfo{author}{Elvira Gamiz}, \bibinfo{author}{Matthias Jamin},
  \bibinfo{author}{Antonio Pich}, \bibinfo{author}{Joaquim Prades} and
  \bibinfo{author}{Felix Schwab}, {\em \bibinfo{journal}{Phys.Rev.Lett.}\/}
  {\bf \bibinfo{volume}{94}}, \bibinfo{pages}{011803} (\bibinfo{year}{2005}).
  \eprint{hep-ph/0408044}.

\bibitem{Barate:1999hj}
\bibinfo{author}{R.~Barate} {\em et~al.\/} (\bibinfo{collaboration}{ALEPH
  Collaboration}), {\em \bibinfo{journal}{Eur.Phys.J.}\/} {\bf
  \bibinfo{volume}{C11}}, \bibinfo{pages}{599} (\bibinfo{year}{1999}).
  \eprint{hep-ex/9903015}.

\bibitem{Chetyrkin:1993hi}
\bibinfo{author}{K.G. Chetyrkin} and \bibinfo{author}{A.~Kwiatkowski}, {\em
  \bibinfo{journal}{Z.Phys.}\/} {\bf \bibinfo{volume}{C59}},
  \bibinfo{pages}{525} (\bibinfo{year}{1993}). \eprint{hep-ph/9805232}.

\bibitem{Maltman:1998qz}
\bibinfo{author}{Kim Maltman}, {\em \bibinfo{journal}{Phys.Rev.}\/} {\bf
  \bibinfo{volume}{D58}}, \bibinfo{pages}{093015} (\bibinfo{year}{1998}).
  \eprint{hep-ph/9804298}.

\bibitem{ckmfitter-vus}
 \bibinfo{note}{V. Niess (CKMfitter Group), Talk at EPS 2011, Grenoble,
  France}, \urlprefix\url{http://ckmfitter.in2p3.fr}.

\bibitem{review}
 \bibinfo{note}{This discussion follows in large parts the
  Review~\cite{hoeckermarciano}.}

\bibitem{Czarnecki:2001pv}
\bibinfo{author}{A.~Czarnecki} and \bibinfo{author}{W.J. Marciano}, {\em
  \bibinfo{journal}{Phys.Rev.}\/} {\bf \bibinfo{volume}{D64}},
  \bibinfo{pages}{013014} (\bibinfo{year}{2001}). \eprint{hep-ph/0102122}.

\bibitem{Davier:2004gb}
\bibinfo{author}{M.~Davier} and \bibinfo{author}{W.J. Marciano}, {\em
  \bibinfo{journal}{Ann.Rev.Nucl.Part.Sci.}\/} {\bf \bibinfo{volume}{54}},
  \bibinfo{pages}{115} (\bibinfo{year}{2004}).

\bibitem{Hanneke:2008tm}
\bibinfo{author}{D.~Hanneke}, \bibinfo{author}{S.~Fogwell} and
  \bibinfo{author}{G.~Gabrielse}, {\em \bibinfo{journal}{Phys.Rev.Lett.}\/}
  {\bf \bibinfo{volume}{100}}, \bibinfo{pages}{120801} (\bibinfo{year}{2008}).
  \eprint{0801.1134}.

\bibitem{Clade:2006zz}
\bibinfo{author}{P.~Clad\'e} {\em et~al.\/}, {\em \bibinfo{journal}{Phys. Rev.
  Lett.}\/} {\bf \bibinfo{volume}{96}}, \bibinfo{pages}{033001}
  (\bibinfo{year}{2006}).

\bibitem{Cadoret:2008st}
\bibinfo{author}{Malo Cadoret} {\em et~al.\/}, {\em \bibinfo{journal}{Phys.
  Rev. Lett.}\/} {\bf \bibinfo{volume}{101}}, \bibinfo{pages}{230801}
  (\bibinfo{year}{2008}). \eprint{0810.3152}.

\bibitem{PhysRevA.74.052109}
\bibinfo{author}{P.~Clad\'e} {\em et~al.\/}, {\em \bibinfo{journal}{Phys. Rev.
  A.}\/} {\bf \bibinfo{volume}{74}}, \bibinfo{pages}{052109}
  (\bibinfo{year}{2006}).

\bibitem{Gerginov:2006zz}
\bibinfo{author}{V.~Gerginov} {\em et~al.\/}, {\em \bibinfo{journal}{Phys.
  Rev.}\/} {\bf \bibinfo{volume}{A73}}, \bibinfo{pages}{032504}
  (\bibinfo{year}{2006}).

\bibitem{Miller:2007kk}
\bibinfo{author}{J.P. Miller}, \bibinfo{author}{E.~de~Rafael} and
  \bibinfo{author}{B.L. Roberts}, {\em \bibinfo{journal}{Rept.Prog.Phys.}\/}
  {\bf \bibinfo{volume}{70}}, \bibinfo{pages}{795} (\bibinfo{year}{2007}).
  \eprint{hep-ph/0703049}.

\bibitem{Jegerlehner:2009ry}
\bibinfo{author}{F.~Jegerlehner} and \bibinfo{author}{A.~Nyffeler}, {\em
  \bibinfo{journal}{Phys.Rept.}\/} {\bf \bibinfo{volume}{477}},
  \bibinfo{pages}{1} (\bibinfo{year}{2009}). \eprint{0902.3360}.

\bibitem{Bennett:2002jb}
\bibinfo{author}{G.W. Bennett} {\em et~al.\/} (\bibinfo{collaboration}{Muon
  g-2}), {\em \bibinfo{journal}{Phys. Rev. Lett.}\/} {\bf
  \bibinfo{volume}{89}}, \bibinfo{pages}{101804} (\bibinfo{year}{2002}).
  \bibinfo{note}{[Erratum-ibid.89:129903,2002]}, \eprint{hep-ex/0208001}.

\bibitem{Bennett:2004pv}
\bibinfo{author}{G.W. Bennett} {\em et~al.\/} (\bibinfo{collaboration}{Muon
  g-2}), {\em \bibinfo{journal}{Phys. Rev. Lett.}\/} {\bf
  \bibinfo{volume}{92}}, \bibinfo{pages}{161802} (\bibinfo{year}{2004}).
  \eprint{hep-ex/0401008}.

\bibitem{Bennett:2006fi}
\bibinfo{author}{G.W. Bennett} {\em et~al.\/} (\bibinfo{collaboration}{Muon
  G-2}), {\em \bibinfo{journal}{Phys. Rev.}\/} {\bf \bibinfo{volume}{D73}},
  \bibinfo{pages}{072003} (\bibinfo{year}{2006}). \eprint{hep-ex/0602035}.

\bibitem{Bailey:1978mn}
\bibinfo{author}{J.~Bailey} {\em et~al.\/}
  (\bibinfo{collaboration}{CERN-Mainz-Daresbury}), {\em \bibinfo{journal}{Nucl.
  Phys.}\/} {\bf \bibinfo{volume}{B150}}, \bibinfo{pages}{1}
  (\bibinfo{year}{1979}).

\bibitem{Schwinger:1948iu}
\bibinfo{author}{Julian~S. Schwinger}, {\em \bibinfo{journal}{Phys.Rev.}\/}
  {\bf \bibinfo{volume}{73}}, \bibinfo{pages}{416} (\bibinfo{year}{1948}).

\bibitem{Kinoshita:2005zr}
\bibinfo{author}{T.~Kinoshita} and \bibinfo{author}{M.~Nio}, {\em
  \bibinfo{journal}{Phys. Rev.}\/} {\bf \bibinfo{volume}{D73}},
  \bibinfo{pages}{013003} (\bibinfo{year}{2006}). \eprint{hep-ph/0507249}.

\bibitem{Aoyama:2007dv}
\bibinfo{author}{T.~Aoyama}, \bibinfo{author}{M.~Hayakawa},
  \bibinfo{author}{T.~Kinoshita} and \bibinfo{author}{M.~Nio}, {\em
  \bibinfo{journal}{Phys. Rev. Lett.}\/} {\bf \bibinfo{volume}{99}},
  \bibinfo{pages}{110406} (\bibinfo{year}{2007}). \eprint{0706.3496}.

\bibitem{Kinoshita:2004wi}
\bibinfo{author}{T.~Kinoshita} and \bibinfo{author}{M.~Nio}, {\em
  \bibinfo{journal}{Phys. Rev.}\/} {\bf \bibinfo{volume}{D70}},
  \bibinfo{pages}{113001} (\bibinfo{year}{2004}). \eprint{hep-ph/0402206}.

\bibitem{Kinoshita:2005sm}
\bibinfo{author}{T.~Kinoshita} and \bibinfo{author}{M.~Nio}, {\em
  \bibinfo{journal}{Phys. Rev.}\/} {\bf \bibinfo{volume}{D73}},
  \bibinfo{pages}{053007} (\bibinfo{year}{2006}). \eprint{hep-ph/0512330}.

\bibitem{Kataev:2006gx}
\bibinfo{author}{A.L. Kataev}  (\bibinfo{year}{2006}). \eprint{hep-ph/0602098}.

\bibitem{Kataev:2006yh}
\bibinfo{author}{A.~L. Kataev}, {\em \bibinfo{journal}{Phys. Rev.}\/} {\bf
  \bibinfo{volume}{D74}}, \bibinfo{pages}{073011} (\bibinfo{year}{2006}).
  \eprint{hep-ph/0608120}.

\bibitem{Passera:2004bj}
\bibinfo{author}{M.~Passera}, {\em \bibinfo{journal}{J. Phys.}\/} {\bf
  \bibinfo{volume}{G31}}, \bibinfo{pages}{R75} (\bibinfo{year}{2005}).
  \eprint{hep-ph/0411168}.

\bibitem{Gabrielse:2006gg}
\bibinfo{author}{G.~Gabrielse}, \bibinfo{author}{D.~Hanneke},
  \bibinfo{author}{T.~Kinoshita}, \bibinfo{author}{M.~Nio} and
  \bibinfo{author}{B.C. Odom}, {\em \bibinfo{journal}{Phys. Rev. Lett.}\/} {\bf
  \bibinfo{volume}{97}}, \bibinfo{pages}{030802} (\bibinfo{year}{2006}).
  \bibinfo{note}{[Erratum-ibid.99:039902,2007]}.

\bibitem{Jackiw:1972jz}
\bibinfo{author}{R.~Jackiw} and \bibinfo{author}{S.~Weinberg}, {\em
  \bibinfo{journal}{Phys. Rev.}\/} {\bf \bibinfo{volume}{D5}},
  \bibinfo{pages}{2396} (\bibinfo{year}{1972}).

\bibitem{Czarnecki:2002nt}
\bibinfo{author}{A.~Czarnecki}, \bibinfo{author}{W.J. Marciano} and
  \bibinfo{author}{A.~Vainshtein}, {\em \bibinfo{journal}{Phys. Rev.}\/} {\bf
  \bibinfo{volume}{D67}}, \bibinfo{pages}{073006} (\bibinfo{year}{2003}).
  \bibinfo{note}{[Erratum-ibid.D73:119901,2006]}, \eprint{hep-ph/0212229}.

\bibitem{Heinemeyer:2004yq}
\bibinfo{author}{S.~Heinemeyer}, \bibinfo{author}{D.~Stockinger} and
  \bibinfo{author}{G.~Weiglein}, {\em \bibinfo{journal}{Nucl. Phys.}\/} {\bf
  \bibinfo{volume}{B699}}, \bibinfo{pages}{103} (\bibinfo{year}{2004}).
  \eprint{hep-ph/0405255}.

\bibitem{Gribouk:2005ee}
\bibinfo{author}{T.~Gribouk} and \bibinfo{author}{A.~Czarnecki}, {\em
  \bibinfo{journal}{Phys. Rev.}\/} {\bf \bibinfo{volume}{D72}},
  \bibinfo{pages}{053016} (\bibinfo{year}{2005}). \eprint{hep-ph/0509205}.

\bibitem{Czarnecki:1995sz}
\bibinfo{author}{A.~Czarnecki}, \bibinfo{author}{B.~Krause} and
  \bibinfo{author}{W.J. Marciano}, {\em \bibinfo{journal}{Phys. Rev. Lett.}\/}
  {\bf \bibinfo{volume}{76}}, \bibinfo{pages}{3267} (\bibinfo{year}{1996}).
  \eprint{hep-ph/9512369}.

\bibitem{Czarnecki:1995wq}
\bibinfo{author}{A.~Czarnecki}, \bibinfo{author}{B.~Krause} and
  \bibinfo{author}{W.J. Marciano}, {\em \bibinfo{journal}{Phys. Rev.}\/} {\bf
  \bibinfo{volume}{D52}}, \bibinfo{pages}{2619} (\bibinfo{year}{1995}).
  \eprint{hep-ph/9506256}.

\bibitem{Peris:1995bb}
\bibinfo{author}{S.~Peris}, \bibinfo{author}{M.~Perrottet} and
  \bibinfo{author}{E.~de~Rafael}, {\em \bibinfo{journal}{Phys. Lett.}\/} {\bf
  \bibinfo{volume}{B355}}, \bibinfo{pages}{523} (\bibinfo{year}{1995}).
  \eprint{hep-ph/9505405}.

\bibitem{Kukhto:1992qv}
\bibinfo{author}{T.V. Kukhto}, \bibinfo{author}{E.A. Kuraev},
  \bibinfo{author}{Z.K. Silagadze} and \bibinfo{author}{A.~Schiller}, {\em
  \bibinfo{journal}{Nucl. Phys.}\/} {\bf \bibinfo{volume}{B371}},
  \bibinfo{pages}{567} (\bibinfo{year}{1992}).

\bibitem{Degrassi:1998es}
\bibinfo{author}{G.~Degrassi} and \bibinfo{author}{G.F. Giudice}, {\em
  \bibinfo{journal}{Phys. Rev.}\/} {\bf \bibinfo{volume}{D58}},
  \bibinfo{pages}{053007} (\bibinfo{year}{1998}). \eprint{hep-ph/9803384}.

\bibitem{Feng:2011zk}
\bibinfo{author}{Xu~Feng}, \bibinfo{author}{Karl Jansen},
  \bibinfo{author}{Marcus Petschlies} and \bibinfo{author}{Dru~B. Renner}
  (\bibinfo{year}{2011}). \eprint{1103.4818}.

\bibitem{Bouchiat:1957zz}
\bibinfo{author}{C.~Bouchiat} and \bibinfo{author}{L.~Michel}, {\em
  \bibinfo{journal}{Phys.Rev.}\/} {\bf \bibinfo{volume}{106}},
  \bibinfo{pages}{170} (\bibinfo{year}{1957}).

\bibitem{Gourdin:1969dm}
\bibinfo{author}{M.~Gourdin} and \bibinfo{author}{E.~De~Rafael}, {\em
  \bibinfo{journal}{Nucl.Phys.}\/} {\bf \bibinfo{volume}{B10}},
  \bibinfo{pages}{667} (\bibinfo{year}{1969}).

\bibitem{Brodsky:1967sr}
\bibinfo{author}{S.J. Brodsky} and \bibinfo{author}{E.~De~Rafael}, {\em
  \bibinfo{journal}{Phys.Rev.}\/} {\bf \bibinfo{volume}{168}},
  \bibinfo{pages}{1620} (\bibinfo{year}{1968}).

\bibitem{:2009fg}
\bibinfo{author}{BABAR Collaboration}, {\em
  \bibinfo{journal}{Phys.Rev.Lett.}\/} {\bf \bibinfo{volume}{103}},
  \bibinfo{pages}{231801} (\bibinfo{year}{2009}). \eprint{0908.3589}.

\bibitem{Arbuzov:1998te}
\bibinfo{author}{A.B. Arbuzov}, \bibinfo{author}{E.A. Kuraev},
  \bibinfo{author}{N.P. Merenkov} and \bibinfo{author}{L.~Trentadue}, {\em
  \bibinfo{journal}{JHEP}\/} {\bf \bibinfo{volume}{9812}}, \bibinfo{pages}{009}
  (\bibinfo{year}{1998}). \eprint{hep-ph/9804430}.

\bibitem{Binner:1999bt}
\bibinfo{author}{S.~Binner}, \bibinfo{author}{J.H. Kuhn} and
  \bibinfo{author}{K.~Melnikov}, {\em \bibinfo{journal}{Phys.Lett.}\/} {\bf
  \bibinfo{volume}{B459}}, \bibinfo{pages}{279} (\bibinfo{year}{1999}).
  \eprint{hep-ph/9902399}.

\bibitem{Davier:2010nc}
\bibinfo{author}{M.~Davier}, \bibinfo{author}{A.~Hoecker},
  \bibinfo{author}{B.~Malaescu} and \bibinfo{author}{Z.~Zhang}, {\em
  \bibinfo{journal}{Eur.Phys.J.}\/} {\bf \bibinfo{volume}{C71}},
  \bibinfo{pages}{1515} (\bibinfo{year}{2011}). \eprint{1010.4180}.

\bibitem{Ambrosino:2010bv}
\bibinfo{author}{KLOE Collaboration}, {\em \bibinfo{journal}{Phys.Lett.}\/}
  {\bf \bibinfo{volume}{B700}}, \bibinfo{pages}{102} (\bibinfo{year}{2011}).
  \eprint{1006.5313}.

\bibitem{:2008en}
\bibinfo{author}{KLOE Collaboration}, {\em \bibinfo{journal}{Phys.Lett.}\/}
  {\bf \bibinfo{volume}{B670}}, \bibinfo{pages}{285} (\bibinfo{year}{2009}).
  \eprint{0809.3950}.

\bibitem{kloeeps}
 \bibinfo{note}{G. Venanzoni (KLOE Collaboration), Talk at EPS-HEP 2011,
  Grenoble, France}.

\bibitem{babar-2pi2pi0}
 \bibinfo{note}{V.P.~Druzhinin (BABAR Collaboration), Talk at the $23^{\rm rd}$
  International Symposium on Lepton-Photon Interactions at High Energy (LP07),
  Daegu, Korea, 13-18 Aug 2007, published in Daegu 2007, Lepton and photon
  interactions at high energies 134, arXiv:0710.3455.}

\bibitem{Aubert:2005eg}
\bibinfo{author}{BABAR Collaboration}, {\em \bibinfo{journal}{Phys.Rev.}\/}
  {\bf \bibinfo{volume}{D71}}, \bibinfo{pages}{052001} (\bibinfo{year}{2005}).
  \eprint{hep-ex/0502025}.

\bibitem{Alemany:1997tn}
\bibinfo{author}{R.~Alemany}, \bibinfo{author}{M.~Davier} and
  \bibinfo{author}{A.~Hoecker}, {\em \bibinfo{journal}{Eur.Phys.J.}\/} {\bf
  \bibinfo{volume}{C2}}, \bibinfo{pages}{123} (\bibinfo{year}{1998}).
  \eprint{hep-ph/9703220}.

\bibitem{Davier:2009ag}
\bibinfo{author}{M.~Davier}, \bibinfo{author}{A.~Hoecker},
  \bibinfo{author}{G.~Lopez~Castro}, \bibinfo{author}{B.~Malaescu},
  \bibinfo{author}{X.H. Mo} {\em et~al.\/}, {\em
  \bibinfo{journal}{Eur.Phys.J.}\/} {\bf \bibinfo{volume}{C66}},
  \bibinfo{pages}{127} (\bibinfo{year}{2010}). \eprint{0906.5443}.

\bibitem{Hagiwara:2011af}
\bibinfo{author}{K.~Hagiwara}, \bibinfo{author}{R.~Liao}, \bibinfo{author}{A.D.
  Martin}, \bibinfo{author}{D.~Nomura} and \bibinfo{author}{T.~Teubner}, {\em
  \bibinfo{journal}{J.Phys.G}\/} {\bf \bibinfo{volume}{G38}},
  \bibinfo{pages}{085003} (\bibinfo{year}{2011}). \eprint{1105.3149}.

\bibitem{Krause:1996rf}
\bibinfo{author}{B.~Krause}, {\em \bibinfo{journal}{Phys.Lett.}\/} {\bf
  \bibinfo{volume}{B390}}, \bibinfo{pages}{392} (\bibinfo{year}{1997}).
  \eprint{hep-ph/9607259}.

\bibitem{lbls}
\bibinfo{note}{Some representative recent estimates of the hadronic
  light-by-light scattering contribution, $\amuhadHOLBLS$, that followed after
  the sign correction of~\cite{Knecht:2001qf,Knecht:2001qg}, are:
  $(105\pm26)\cdot10^{-11}$~\cite{Prades:2009tw},
  $(110\pm40)\cdot10^{-11}$~\cite{Bijnens:2007pz},
  $(136\pm25)\cdot10^{-11}$~\cite{Melnikov:2003xd}.}

\bibitem{Bijnens:2007pz}
\bibinfo{author}{J.~Bijnens} and \bibinfo{author}{J.~Prades}, {\em
  \bibinfo{journal}{Mod.Phys.Lett.}\/} {\bf \bibinfo{volume}{A22}},
  \bibinfo{pages}{767} (\bibinfo{year}{2007}). \eprint{hep-ph/0702170}.

\bibitem{Prades:2009tw}
\bibinfo{author}{J.~Prades}, \bibinfo{author}{E.~de~Rafael} and
  \bibinfo{author}{A.~Vainshtein}  (\bibinfo{year}{2009}). \eprint{0901.0306}.

\bibitem{Melnikov:2003xd}
\bibinfo{author}{K.~Melnikov} and \bibinfo{author}{A.~Vainshtein}, {\em
  \bibinfo{journal}{Phys.Rev.}\/} {\bf \bibinfo{volume}{D70}},
  \bibinfo{pages}{113006} (\bibinfo{year}{2004}). \eprint{hep-ph/0312226}.

\bibitem{deRafael:1993za}
\bibinfo{author}{E.~de~Rafael}, {\em \bibinfo{journal}{Phys.Lett.}\/} {\bf
  \bibinfo{volume}{B322}}, \bibinfo{pages}{239} (\bibinfo{year}{1994}).
  \eprint{hep-ph/9311316}.

\bibitem{hoeckermarciano}
\bibinfo{author}{A.~Hoecker} and \bibinfo{author}{W.~Marciano}
  \bibinfo{note}{{\em The Muon Anomalous Magnetic Moment}, in: Particle Data
  Group (K.~Nakamura \ea), J. Phys. G 37, 075021 (2010)}.

\bibitem{Aad:2011ib}
\bibinfo{author}{ATLAS Collaboration}  (\bibinfo{year}{2011}).
  \eprint{1109.6572}.

\bibitem{Chatrchyan:2011zy}
\bibinfo{author}{CMS Collaboration}, {\em \bibinfo{journal}{Phys.Rev.Lett.}\/}
  {\bf \bibinfo{volume}{107}}, \bibinfo{pages}{221804} (\bibinfo{year}{2011}).
  \eprint{1109.2352}.

\bibitem{Buchmueller:2011ki}
\bibinfo{author}{O.~Buchmueller}, \bibinfo{author}{R.~Cavanaugh},
  \bibinfo{author}{D.~Colling}, \bibinfo{author}{A.~De~Roeck},
  \bibinfo{author}{M.J. Dolan} {\em et~al.\/}, {\em
  \bibinfo{journal}{Eur.Phys.J.}\/} {\bf \bibinfo{volume}{C71}},
  \bibinfo{pages}{1722} (\bibinfo{year}{2011}). \eprint{1106.2529}.

\bibitem{Pospelov:2008zw}
\bibinfo{author}{M.~Pospelov}, {\em \bibinfo{journal}{Phys.Rev.}\/} {\bf
  \bibinfo{volume}{D80}}, \bibinfo{pages}{095002} (\bibinfo{year}{2009}).
  \eprint{0811.1030}.

\bibitem{TuckerSmith:2010ra}
\bibinfo{author}{D.~Tucker-Smith} and \bibinfo{author}{I.~Yavin}, {\em
  \bibinfo{journal}{Phys.Rev.}\/} {\bf \bibinfo{volume}{D83}},
  \bibinfo{pages}{101702} (\bibinfo{year}{2011}). \eprint{1011.4922}.

\bibitem{fnal_g-2}
 \bibinfo{note}{New $g-2$ Collaboration},
  \urlprefix\url{{http://gm2.fnal.gov/public_docs/proposals/Proposal-APR5-Final.pdf}}.

\bibitem{jparc_g-2}
 \bibinfo{note}{V. Vrba {\em et al.}, Report KEK\_J-PARC-PAC2009-06, See also,
  \eg, Naohito SAITO (KEK), Seminar at DESY 2011}.

\bibitem{Knecht:2001qf}
\bibinfo{author}{M.~Knecht} and \bibinfo{author}{A.~Nyffeler}, {\em
  \bibinfo{journal}{Phys.Rev.}\/} {\bf \bibinfo{volume}{D65}},
  \bibinfo{pages}{073034} (\bibinfo{year}{2002}). \eprint{hep-ph/0111058}.

\bibitem{Knecht:2001qg}
\bibinfo{author}{M.~Knecht}, \bibinfo{author}{A.~Nyffeler},
  \bibinfo{author}{M.~Perrottet} and \bibinfo{author}{E.~de~Rafael}, {\em
  \bibinfo{journal}{Phys.Rev.Lett.}\/} {\bf \bibinfo{volume}{88}},
  \bibinfo{pages}{071802} (\bibinfo{year}{2002}). \eprint{hep-ph/0111059}.

\end{thebibliography}

\end{document}